\documentclass[aps,showpacs,eqsecnum,twocolumn,superscriptaddress,floatfix]
{revtex4}

\usepackage{amsmath}
\usepackage{latexsym}
\usepackage{times}
\usepackage{amssymb}
\usepackage{color}

\usepackage{graphicx}

\newcommand{\be}{\begin{equation}}
\newcommand{\ee}{\end{equation}}
\newcommand{\bea}{\begin{eqnarray}}
\newcommand{\eea}{\end{eqnarray}}
\newcommand{\no}{\noindent}
\newcommand{\nn}{\nonumber}

\newcommand{\Tr}{{\rm Tr\,}}
\newcommand{\Det}{{\rm Det\,}}
\newcommand{\e}{{\rm e\,}}

\newcommand{\parm}{\par\medskip}

\newcommand{\rar}{\rightarrow}

\newcommand{\C}{\mathbb{C}}

\newcommand{\one}{\mbox{1\hspace{-.8ex}I}}

\begin{document}

\title[A Model for QCD at High Density and Large Quark Mass]{
A Model for QCD at High Density and Large Quark Mass}

\author{Roberto De Pietri}
\affiliation{Dipartimento di Fisica, Universit\`a di Parma, Italy} 
\affiliation{INFN Gruppo Collegato di Parma, Italy}
\author{Alessandra Feo}
\affiliation{Dipartimento di Fisica, Universit\`a di Parma, Italy} 
\affiliation{INFN Gruppo Collegato di Parma, Italy}
\author{Erhard Seiler}
\affiliation{
Max-Planck-Institut f\"ur Physik (Werner Heisenberg Institut),
M\"unchen, Germany}
\author{Ion-Olimpiu Stamatescu}
\affiliation{FEST - Protestant Institute for Interdisciplinary Research, Heidelberg, Germany}
\affiliation{Institut f\"ur Theoretische Physik  der Universit\"at, Heidelberg, Germany}

\date{\today} \begin{abstract} 
We study the high density region of QCD within an effective model obtained 
in the frame of the hopping parameter expansion and choosing Polyakov
type of loops as the main dynamical variables representing
the fermionic matter. To get a first idea of 
the phase structure, the model is analyzed in strong
coupling expansion and using a mean field 
approximation. In numerical simulations, the model still shows the 
so-called sign problem, a difficulty peculiar to non-zero chemical 
potential, but it permits the development of algorithms which ensure a 
good overlap of the Monte Carlo ensemble with the true one. We review the 
main features of the model
and present calculations concerning the dependence of 
various observables on the chemical potential and on the temperature, in 
particular of the charge density and the diquark susceptibility, which may 
be used to characterize the various phases expected at high baryonic 
density. We obtain in this way information about the phase structure of the
model and the corresponding phase transitions and cross over regions, which
can be considered as hints for the behaviour of non-zero density QCD.
\end{abstract}

\maketitle

\section{Introduction}
The exploration of the phase diagram of matter at non-zero baryon 
density is a challenging and interesting problem. 
In particular, it has
been emphasized that quark matter at extremely high density may 
behave as a color superconductor (see Ref.\cite{alford} for a 
recent review on the subject and references therein).
Moreover, it is also expected that the phase diagram in the 
temperature-density plane shows multiple phases separated 
by various critical lines and, except for the high $T$, small $\mu$ region, 
not much is known about their exact position and nature. 

Lattice gauge theory calculations in various implementations  
that try to evade the sign problem generated by the non-zero
chemical potential have 
been mostly performed at small baryon density and high temperature,
where they agree reasonably well with each other. Here there is good
evidence for the presence of a crossover instead of a sharp deconfining
transition.
At large $\mu$ 
(baryon density), however, there are only few numerical results 
which need to be corroborated by using different methods. See
 \cite{karrev} for a review.

The aim of this work is to understand the phase structure of high
density, strongly interacting matter. 
Most work on QCD at non-zero density proceeds from the $\mu=0$,
$T\sim T_c$ region and attempts to go as far as possible in the
$\mu > 0$ domain. As an alternative one may consider the possibility
to start from the large $\mu$ domain and try to reach the region of
interest from above.
In the spirit of the $\mu=0$
quenched approximation a `non-zero density
quenched approximation' for $\mu > 0$ based on the double limit $ M
\rar \infty,\, \mu \rar \infty,\, \zeta \equiv {\rm exp}\,(\mu - \ln
M) :$ fixed \cite{bend,fktre} has been considered. This implements a 
static, charged
background, which influences the gluonic dynamics
\cite{fktre,bky}. The present model \cite{hdm01} represents a
systematic extension of the above considerations: the gluonic vacuum
is enriched by the effects of dynamical quarks of large (but not
infinite) mass, providing a large net baryonic charge.  In \cite{hs}
and in the present paper we explore the phase structure of the model,
as a first step in understanding the properties of such a background.
 
This model can be derived as a $1/M$ expansion of QCD at large $\mu$
around the unphysical limit of infinitely heavy quarks. 
However, it is more
realistic to understand it as an approximation whose justification
relies on the predominant role of the gluonic dynamics. We want to
understand how this dynamics is influenced by the presence of charged
matter. This would allow, among other things, to study the effect of
dense, heavier background baryonic charges on light quarks and hadrons.

The main ingredient of the model are Polyakov-type loops,
capturing the effect of heavy quarks with low mobility. The model
still has a sign problem, but being based on the variables which are 
especially sensitive to the physics of dense baryonic matter it 
allows for reweighting
algorithms which ensure a good overlap 
of the Monte Carlo ensemble with the true one. 

The paper is organized as follows. 
In Sec.\ref{model} we study the high density region of QCD
within an effective model obtained by an expansion in the hopping 
parameter $\kappa$ of the fermionic determinant up to next-to leading order, 
$\kappa^2$. In Sec.\ref{analytic} the model is analyzed using first a 
strong coupling expansion and then a mean field 
approximation just to get a first idea of the phase diagram 
and to compare with numerical simulations.

Sec.\ref{numeric} shows results of the numerical simulations.
Here the model shows the so-called sign problem 
but due to the factorization of the
fermionic determinant it  permits to develop 
very efficient local algorithms and
achieve large statistics.
The dependence of various observables
on the chemical potential and the temperature is studied
and we show a tentative phase diagram at 
large mass and
high baryon density. 
Conclusions and outlook are given in Sec.~\ref{conclusions}.

\section{QCD at large chemical potential}.
\label{model}

\subsection{QCD at non-zero $\mu$}

In this study we use the grand canonical formulation of QCD, i.e.,
we introduce the chemical potential $\mu$ as a (bare) parameter.
The QCD grand canonical partition function with Wilson fermions at $\mu>0$ 
is:
\begin{align}
& \!\!\!\!\!\!\!\!\!\!\!\!\!\!\!
{\cal Z}(\beta,\kappa,\gamma_G,\gamma_F,\mu) \nn\\
&= \int[DU]\, 
\e^{-S_G(\beta,\gamma_G,\{U\})}{\cal Z}_F({ {\kappa}},\gamma_F, \mu, \{U\}) \, ,
\label{e.gcpt} \\ 
& \!\!\!\!\!\!\!\!\!\!\!\!\!\!\!
S_G(\beta,\gamma_G,\{U\})   \nn \\
&= -\frac{\beta}{N_c}\,Re\,\Tr\,\left(\frac{1}{\gamma_G}\,
\sum_{j>i=1}^3\,
P_{ij} + \gamma_G\,\sum_{i}\,P_{i4}\right)\, , \label{e.YMa}\\ 
& \!\!\!\!\!\!\!\!\!\!\!\!\!\!\!
{\cal Z}_F({ {\kappa}}, \gamma_F,\mu, \{U\}) =  
\Det W ({ {\kappa}}, \gamma_F,\mu, \{U\}) \, , \label{e.det}\\ 
\begin{split} 
W_{ff'} &= \delta_{ff'} [ 1 -  \kappa_f\, \sum_{i=1}^3 \left( 
\Gamma_{+i}\,U_i\,T_i +
\Gamma_{-i}\,T^*_i\,U^*_i\right) \\
& - \kappa_f\, \gamma_F\, \left( \e^{\mu_f}\,\Gamma_{+4}\,U_4\,T_4 +
\e^{-\mu_f}\,\Gamma_{-4}\,T^*_4\,U^*_4 \right) ] \, ,  \\
\Gamma_{\pm \mu} &= 1 \pm \gamma_{\mu},\ \ \gamma_{\mu}=\gamma_{\mu}^*,\
\gamma_{\mu}^2=1 \, , \\ 
\kappa &=
\frac{1}{2(M+3+\gamma_F\,\cosh \mu)} = \frac{1}{2(M_0+3+\gamma_F)} \, ,
\end{split} \nn
\end{align}
\no where we have specialized $S_G$ for Wilson's plaquette ($P$) action 
and used a certain definition of the Wilson term in $W$.
 Here $M$ is the `bare mass', $M_0$  the bare mass at $\mu=0$,
$f$ is the flavor index, $U_\mu$  denote the link variables and $T_\mu$ 
lattice translations.  For the sake of generality and the
discussion in section III.B we also
introduced coupling anisotropies $\gamma_G$, $\gamma_F$ which
 however will be set to 1 
elsewhere. All quantities are understood in units of the (spatial)
lattice spacing $a$ unless explicitly specified otherwise.
The exponential prescription for $\mu$ ensures canceling of 
divergences in the small $a$ limit \cite{hkks}. 
A non-zero physical temperature $T$ is introduced as
\bea
a\,T = \frac{\gamma_{phys}}{N_\tau} \label{e.temp0} \, ,
\eea
where $\gamma_{phys}$ is the physical cutoff anisotropy defined by an 
appropriate renormalization of the coupling anisotropies \cite{bkns},
 and $N_\tau$ the `length'
 of the (periodic) temporal lattice size.

The fermionic coupling matrix $W$ fulfills:
\bea
\gamma_5 W(\mu) \gamma_5 = 
W(-\mu)^*,\ \ \Det W(\mu) = \Det W(-\mu)^*
\eea
where the $*$ conjugation above is understood in the lattice
and color indices, that is $U_{n,\nu}^* =  U^{\dagger}_{(n+\nu),-\nu}$.
At $\mu \ne 0$ the determinant is  complex (while, due
to the symmetries of the Yang-Mills integration the full partition
function remains real). 

Numerical simulations are based on defining an efficient importance 
sampling of the configurations. Since the integrand (for simplicity we shall 
still call it `Boltzmann factor'):
\bea
B = \e^{-S_G(\beta,\{U\})}{\cal Z}_F({ {\kappa}}, \mu, \{U\})
\eea
is not a real, positive definite number it does not define a
probability measure for the Yang-Mills integration. There have been a
number of methods devised to cope with this problem, which all involve
simulating a different ensemble and correcting the results either by
continuing in $\mu$ or by redefining the observables.

Continuation methods use the Taylor expansion \cite{TARO}, \cite{owph}
or more sophisticate expansions \cite{mpl} to enter the region of
real, non-zero $\mu$ by fitting the coefficients from $\mu=0$
simulations \cite{TARO} or from simulations at imaginary $\mu$
\cite{owph} \cite{mpl}.  They rely on correctly identifying the
analytic properties of the partition function and the various
expectation values. Due to the noise in determining the expansion
coefficients the quality of the continuation degrades rapidly with
increasing (real) $\mu$.  Since the simulations are done with
dynamical quarks the statistics is limited.

The so called `reweighting method' proceeds by choosing a 
positive definite measure $B_0$ obtained by splitting the original 
`Boltzmann factor' according to
\bea
B = B_0 w_0 \label{e.b0w0} \, .
\eea
$B_0$ is used to produce an ensemble of configurations $C^0_n= \{U\}^0_n$  
(where $n$ indexes the configurations) to be reweighted by the complex
numbers $w_{0,n} = B_n/B_{0,n}$ associated with the configurations 
$C^0_n$ in calculating expectation values:
\bea
\langle O \rangle = \frac{\langle w_0 O\rangle_0}{\langle w_0 \rangle_0} \, ,
\label{expw} 
\eea
with $O$ some observable and $\langle \dots \rangle_0$ denoting averages
over the ensemble $C^0$. 
Notice that $w_0$ is both complex and non-local since it comes from
the fermionic determinant. The $\langle \dots \rangle_0$
averages contain therefore alternating contributions with large cancellations
(the `sign problem'). Moreover, the reweighting can correct an underestimated
contribution in the $C^0$ ensemble, but fails if the underestimation is too 
drastic (the `overlap problem'). In both cases the problems are aggravated by 
the non-locality of $w_0$ which makes it difficult to achieve high statistics.
 
Calculations based on various implementations of the reweighting method
\cite{rwm} have been performed mainly at small $\mu$, where
they agree reasonably well with other methods (analytic expansion
\cite{owph}, \cite{mpl}, \cite{ejir}). At large $\mu$,
however, there are only few numerical results yet, mainly
based on only one method  \cite{fod} and corroboration 
by different methods is missing.
 
At large $\mu$ the behaviour of QCD quantities may however be dominated by 
certain factors in the fermionic determinant which lead to a simpler model 
that is actually easier to simulate. In its lowest order this model is 
considered to define what can be called `quenched, non-zero density QCD' 
\cite{fktre}.  The model is based on an analytic expansion of QCD (the 
hopping parameter expansion) and involves the Polyakov loop variables of 
the theory, which in many setups are thought to catch important effects of 
the fermionic matter \cite{pol_loop}. This, and its suitability for 
numerical simulations makes this model interesting for study. Moreover it may 
give us hints for improving the algorithms for the full QCD at non-zero density.

In the next 
subsections we shall recall the hopping parameter expansion and describe 
the model.

\subsection{Hopping parameter expansion of the fermionic determinant} 

The large mass (hopping parameter) expansion of QCD arises from an expansion 
of the logarithm of the fermionic determinant exhibiting only closed loops:
\bea
\Det W &=& {\rm exp} ( \Tr 
\ln W )    \label{e.hopg} \\
 &=&  {\rm exp} \left[-\sum_{l=1}^\infty  \sum_{\left\{{\cal
C}_l\right\}} \sum_{s=1}^\infty ~{{{ (\kappa_f^l g^f_{{\cal C}_l})}^s}\over
s}\,\Tr_{\rm D,C}     {\cal L}_{{\cal C}_l}^s \right] \nn \\
 &=& 
\prod_{l=1}^{\infty} ~\prod_{\left\{{\cal C}_l\right\}} ~\prod_{f}~ 
  \Det_{\rm D,C} \left(\one~-~(\kappa_f)^l g^f_{{\cal C}_l}
{\cal L}_{{\cal C}_l}\right) \, .\nn
\eea
Here ${\cal C}_l$ are distinguishable, 
non-exactly-self-repeating 
closed paths of length $l$ and $s$ is the number of times
a loop ${\cal L}_{{\cal C}_l}$ covers  ${\cal C}_l$. 
With $\lambda$ denoting the links along  ${\cal C}_l$ we have
\bea
{\cal L}_{{\cal C}_l} &=& \left(\prod_{\lambda \in {\cal C}_l}\Gamma_{\lambda}
U_{\lambda}\right)^s\, , \label{e.lp} \\
g^f_{{\cal C}_l} &=& \left(\epsilon \, \e^{\pm N_{\tau}\mu_f}\right)^r\ {\rm 
if}\ 
{\cal C}_l = \text{`Polyakov r-path'} \, , \nn\\
\no &=&1\  \text{otherwise} \, . \label{e.fact}
\eea
The index $D,C$ in (\ref{e.hopg}) 
means that the traces (the determinants) are understood both
over Dirac and color indices.
A `Polyakov  r-path'  closes over the lattice 
in the $\pm 4$  direction with winding number $r$ and
periodic(antiperiodic) b.c. ($\epsilon = +1(-1)$). 
We assume periodic b.c. in the `spatial' directions.
Notice that, since the determinant is a polynomial in $\kappa$ this 
expansion terminates at the order $d N_L N_c n_f$ with $d=2,4$ the dimension, 
$N_L$ the lattice volume, $N_c$ the number of colors and 
$n_f$ the number of flavors. For details see \cite{sdet}.

\subsection{The massive, dense limit of the fermionic determinant} 

The double limit \cite{bend}
\bea
\kappa \rightarrow 0,\, \mu \rightarrow \infty ,\ \ 
\kappa\, \e^{\mu} \equiv \zeta:\ {\rm fixed} \label{e.qul}
\eea
produces a static, dense, charged background on the lattice, and has 
been therefore proposed and studied as a non-zero density quenched 
approximation \cite{fktre,bky}. Note that the pure Yang-Mills limit 
corresponds to $\zeta=0$, which for fixed nonzero $\kappa$ requires 
$\mu\to -\infty$. 

In the limit (\ref{e.qul}) the fermionic determinant simplifies 
considerably, e.g., for 1 flavor we have:
\bea 
{\cal Z}_F^{[0]}(C, 
\left\{U\right\}) = \exp \left[-2 
  \sum_{\left\{{\vec x}\right\}}
\sum_{s=1}^\infty\!\! ~{{{ (\epsilon C)}^s}\over s}~\Tr 
  ({\cal P}_{\vec x})^s \right] \nonumber
\eea
\vspace{-0.5cm}
\bea
=\, \prod_{\left\{{\vec x}\right\}}~ 
  \Det \left(\one~-~\epsilon\,C {\cal P}_{\vec x}\right)^2,
\,\,\,\,\,\, C = (2\, \zeta)^{N_\tau} \label{e.hdl}\,, 
\eea 
where ${\cal P}_{\vec x}$ denotes the Polyakov loop
\bea
{\cal P}_{\vec x}\equiv 
\prod _{t=0}^{N_\tau-1}U_{(\vec x,t),\mu} \label{e.polo}
\eea
and from now on traces and determinants are understood only over
the color indices. For later reference we also define the shortening:
\bea
P \equiv \frac{1}{N_c}\,\Tr {\cal P}\, , \ \ \ 
P^{\ast} \equiv \frac{1}{N_c}\,\Tr {\cal P}^{\dagger} \label{e.pdef}
\eea
(notice the different normalization to (\ref{e.polo}) above).
In the limit (\ref{e.qul}) $\mu$ diverges and  the parameter of the model 
is $\zeta$ (\ref{e.qul}) or the related
$C$ (\ref{e.hdl}) which is directly connected to the
average charge density on
a non-zero temperature lattice:
\bea
\hat n_0 &=& \langle \frac{\partial}{\partial \mu}{\cal Z}_F^{[0]} \rangle 
\simeq   2 C \langle \sum_{\vec x}\Tr {\cal P}_{\vec x} \rangle \, . 
\label{e.chd0}  
\eea
One can study the behavior of various quantities, such as gluonic
correlation functions and correlation functions involving 
light quarks on such a static background, much like in the 
quenched approximation at $\mu=0$. However, effects expected to be
due to the mobility of charges, in particular the possibility of
new phases in dependence on the chemical potential cannot be 
studied here. 

Since this limit is obtained in an analytic expansion, we can 
systematically consider higher order corrections. In the following we 
shall study the model which is obtained at the next order.

\subsection{Large $\mu$ limit in order $\kappa^2$ as
a model for high density QCD}

\begin{figure}[t]
\includegraphics[width=.9\columnwidth]{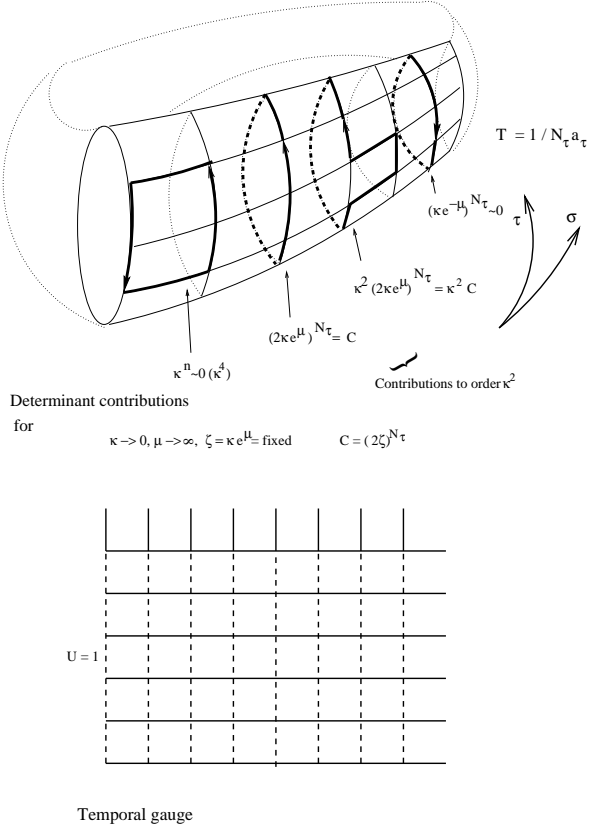}  
\caption{Periodic lattice, loops, temporal gauge. In the maximal temporal gauge
also the links of the basis line are fixed to 1 up to the rightmost one.} 
\label{torus_2} 
\end{figure}

The fermionic determinant to this order is given by:
 \begin{multline}
{\cal Z}_F^{[2]}({ {\kappa}}, \mu, \left\{U\right\}) 
 =   {\rm exp}\left\{-2\,  \sum_{\left\{{\vec x}\right\}}\,
\sum_{s=1}^\infty \,{{{ (\epsilon\, C
)}^s}\over s} \right. \times \\ 
\times \left. \Tr 
  \left[({\cal P}_{\vec x})^s  + \kappa^2\sum_{r,q,i,t,t'}
(\epsilon\, C)^{s(r-1)}({\cal P}_{{\vec x},i,t,t'}^{r,q})^s \right]\right\} \\
= {\cal Z}_F^{[0]}( C,  \left\{U\right\}) 
\prod_{{\vec x}, r,q,i,t,t'} \!\!\!
  \Det \left(\one-(\epsilon\,C)^{r}\,\kappa^2\,
  {\cal P}_{{\vec x},{i},t,t'}^{r,q}\right)^2 . 
\label{e.corr2}
\end{multline}
The loops contributing to the determinant are shown in Fig. \ref{torus_2}.
In the following we shall use antiperiodic b.c. ($\epsilon=-1$) to 
ensure reflection positivity.

For easy bookkeeping we use the temporal gauge
\bea 
U_{n,4}=1,\ {\rm except\ for}\  
U_{({\vec x}, n_4=N_\tau),4} \equiv V_{\vec x}: \ 
{\rm free}\, ,
\eea
then
\bea
{\cal P}_{{\vec x},i,t,t'}^{r,q} &=& (V_{\vec x})^{r-q} 
U_{({\vec x},t),i} (V_{{\vec x}+{\hat {\i}}})^q 
U_{({\vec x},t'),i}^* \\
&&r>q\ge 0,\ i = \pm 1, \pm 2, \pm 3, \nn \\  
&&1\le t \le t'\  \le N_{\tau}\ \ (t < t' \ {\rm for}\ q=0) \, .
\nn
\eea
See \cite{hdm01}. 
Notice that for $SU(3)$ we have:
\bea
 \Det (\one + C\,{\cal P}) &=& 1 + C\, \Tr{\cal P}
+ C^2\,\Tr{\cal P}^* + C^3 \nn \\
&=& 1 + 3C\,P +3C^2\,P^* +C^3\, .
\eea 

Our model is thus defined by using ${\cal Z}_F^{[2]}$ for 
${\cal Z}$ in Eqs.(\ref{e.det},\ref{e.gcpt})
rewritten for general number of flavors $n_f$.
Since ${\cal Z}_F^{[2]}$
is factorizable it is easily calculable. It is suggestive to use 
a splitting Eq. (\ref{e.b0w0}) preserving
 the factorization property 
which would allow to design a local algorithm for producing the 
$C^0$ ensemble. 

Preliminary results have been reported in \cite{hdm01}, \cite{dfss}.
Here we report an extensive analysis of the phase structure of this
model at large $\mu$.

\section{Analytic Computations}
\label{analytic}
\subsection{Strong coupling/hopping parameter expansion}

As a first orientation about the behavior of the model we consider the 
strong coupling and hopping parameter expansion, which will also serve 
as a check of the Monte Carlo results. For simplicity we limit 
ourselves to one flavor here. The expansion proceeds in powers of 
the parameters $\beta$ and $\kappa$; we are mainly interested in the 
results for the expectation values $\langle P_{\vec x}\rangle$ of 
the Polyakov loop and its adjoint $\langle P_{\vec x}^{\ast}\rangle$.

\begin{figure}[htb]
\includegraphics[width=.9\columnwidth]{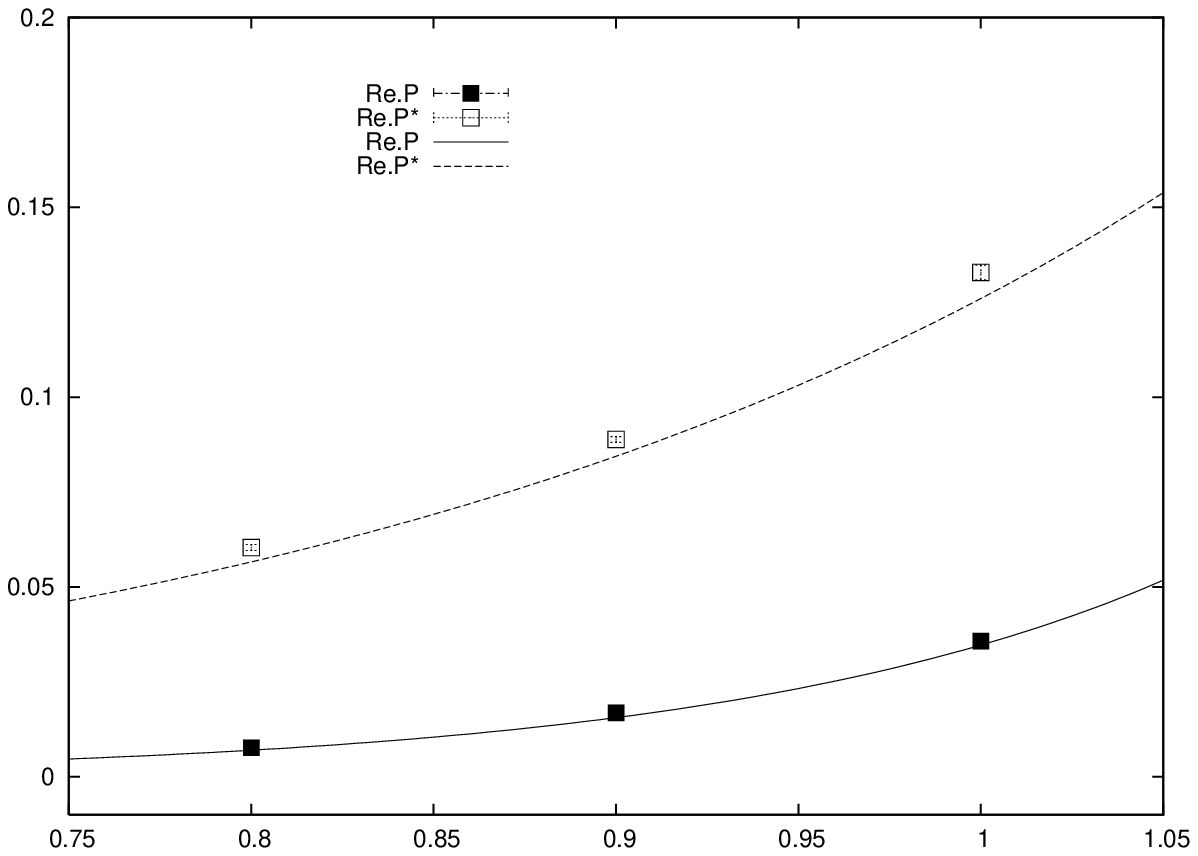}  
\includegraphics[width=.9\columnwidth]{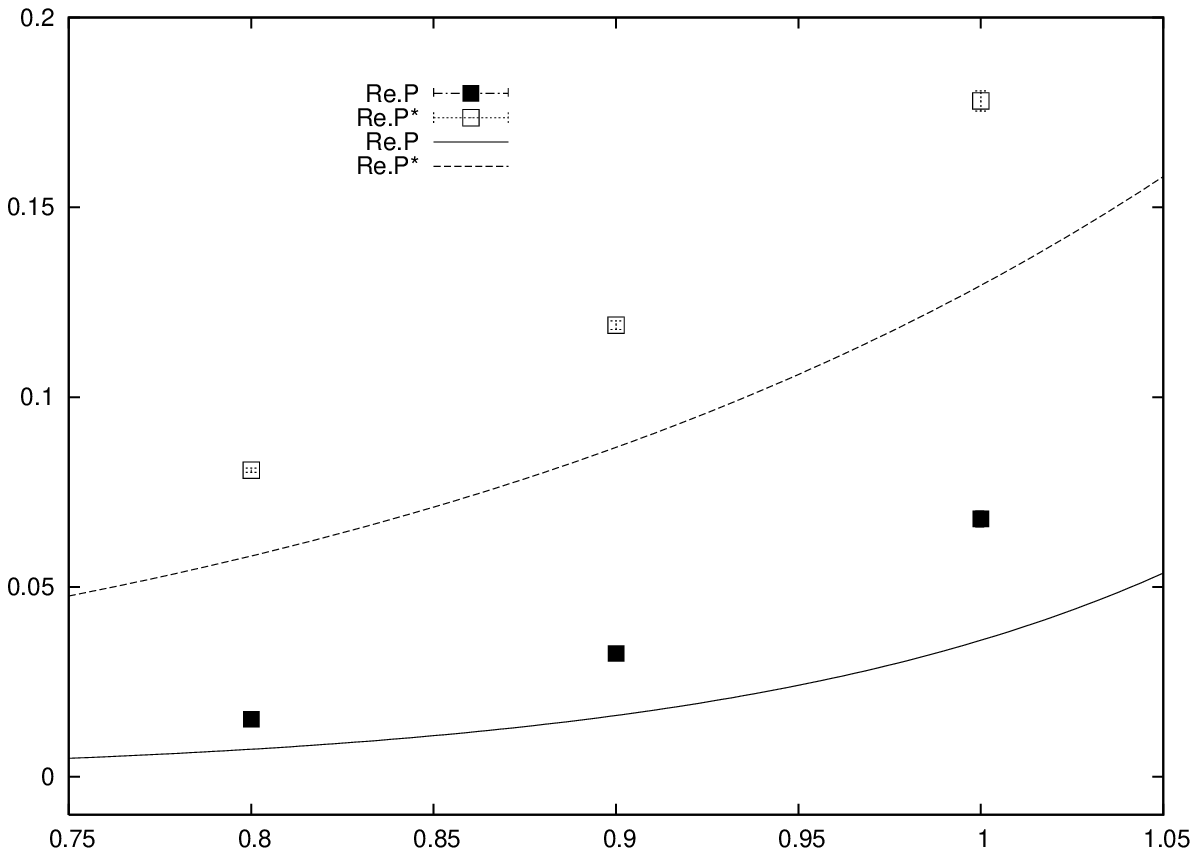}  
\caption{Comparison with strong coupling at $\beta=3$ (upper plot) and
$\beta=5$ (lower plot), $4^4$ lattice.
Full symbols denote $Re P$, empty symbols $Re P^\ast$, the lines show the 
corresponding strong coupling results.} 
\label{sc443} 
\end{figure}

\begin{figure}[htb]
\includegraphics[width=.9\columnwidth]{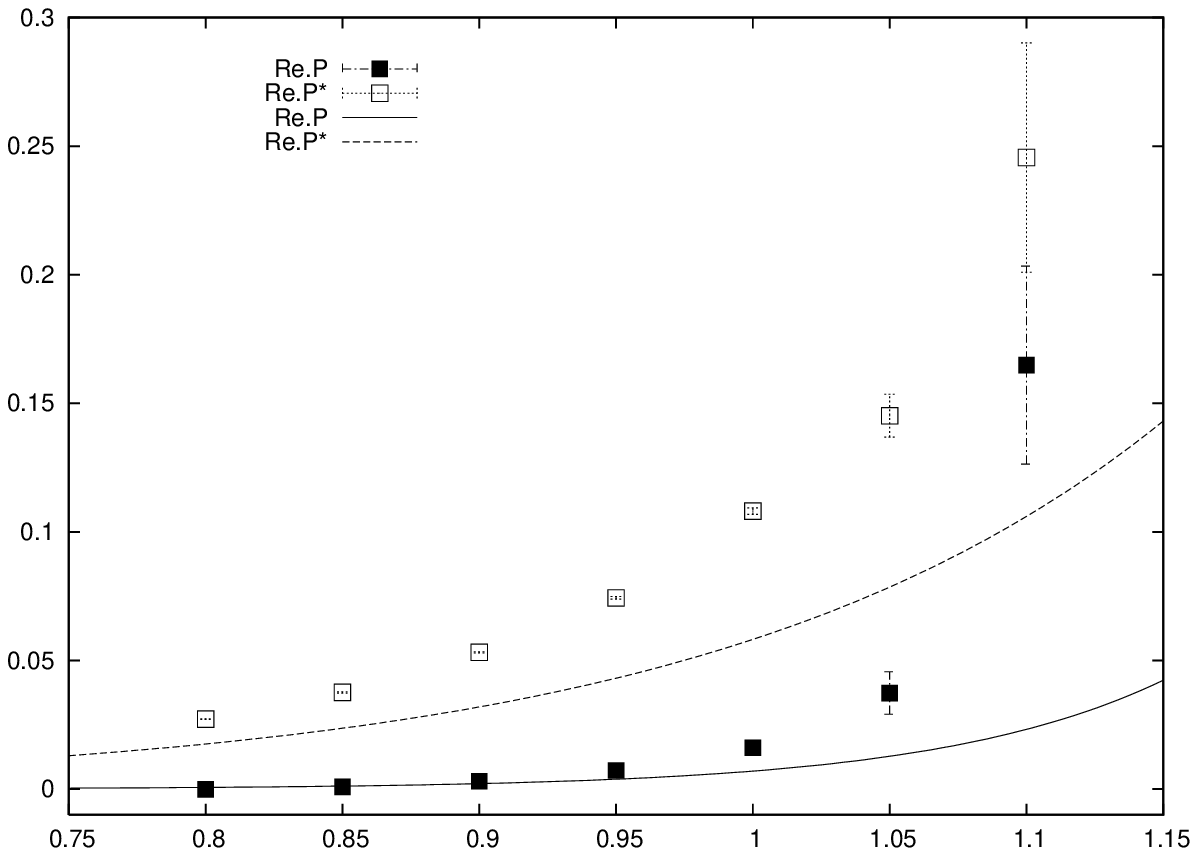}
\includegraphics[width=.9\columnwidth]{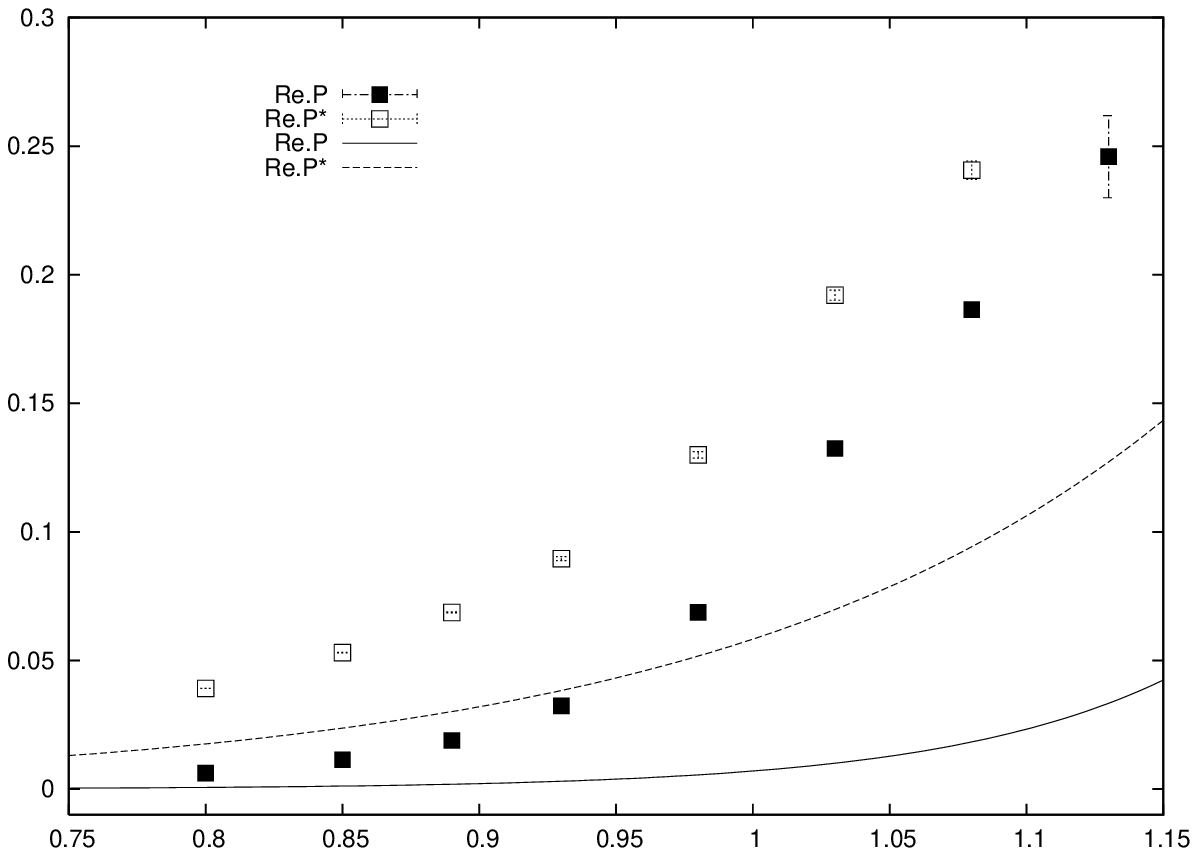}
\caption{Comparison with strong coupling, $\beta=5.5$ (upper plot) and
$\beta=5.6$ (lower plot), $6^4$ lattice.
Symbols as in Fig.\ref{sc443}}
\label{sc6655}
\end{figure}

Some details of the computation are given in Appendix A. The results for 
$\langle P\rangle$ and  $\langle P^\ast \rangle$ to order $\kappa^2$ are
\begin{multline}
\langle P\rangle^{[2]}\equiv C^2 \frac{1+\frac{2}{3} C^3}{1+4C^3+C^6}
\Biggl[1+\cr\frac{2\beta\kappa^2(N_\tau-1)}{3}
\frac{2+3C^2+6C^6}{(1+4C^3+C^6)(3+2C^3)}\Biggr]
\end{multline}
and
\begin{multline}
\langle P^\ast \rangle^{[2]} \equiv 
C \frac{\frac{2}{3}+C^3}{1+4C^3+C^6} 
\Biggl[1+\cr\frac{2\beta\kappa^2(N_\tau-1)}{3}
\frac{(1+C^3)^4+7C^6}{(1+4C^3+C^6)(2+3C^3)}\Biggr]\ .
\end{multline}
The leading behavior of this for small $C$ is
\be
\langle P\rangle^{[2]}\sim 
C^2\left(1+\frac{4}{9}\beta\kappa^2(N_\tau-1)\right) 
\end{equation}
and
\be
\langle P^\ast\rangle^{[2]}\sim \frac{2}{3}C
\left(1+\frac{1}{3}\beta\kappa^2(N_\tau-1)\right)\rangle \ .
\end{equation}
In Figs. \ref{sc443} and \ref{sc6655}  we 
compare the results for $P$ and $P^\ast$ of the Monte Carlo simulations 
on $4^4$ and $6^4$ lattices, for $\kappa=0.12$, one flavor and 
different values of $\beta$, with $P^{[2]}$ and $P^{\ast [2]}$. The 
agreement is good for the $4^4$ lattice and $\beta=3$, 
while for $\beta=5$ there are already significant deviations.
But the agreement between Monte Carlo and strong coupling 
results is sufficient to validate the simulations. 
On the other hand, on the $6^4$ lattice there is a remarkable difference 
between $\beta=5.5$ and $5.6$; while in the former case the agreement with
the strong coupling expansion remains good up to $\mu\approx 0.95$ at least for 
$\langle P\rangle$, in the 
latter case the simulation results start deviating from strong coupling 
at much lower values of $\mu$. This can be seen as an indication of a 
phase transition in this region.

\subsection{Mean field calculations}

Mean field calculations were quite popular in the early years of lattice 
gauge theory. They generally gave reasonably good indications of the phase 
structure of various models, but with the development of high speed 
computers and the corresponding improvement of Monte Carlo calculations 
they fell more or less into oblivion. The reason we are reviving them here 
is to get some qualitative insight into the phase structure of our model 
to which the Monte Carlo simulation can be compared. But it should be kept 
in mind that the method suffers from a certain amount of non-uniqueness 
and one has to apply it with some common sense. Since the mean field 
approximation of our model shows some peculiarities and has not been 
discussed anywhere in the literature, we found it necessary to derive it 
from the beginning. We summarize here the results and give details in the 
appendix.

The experience with mean field theory showed that its quality is poor 
without gauge fixing, but with temporal gauge fixing in pure 
Yang-Mills theory at zero temperature one gets reasonable results. Since 
we are dealing here with finite temperature, temporal gauge fixing is not 
possible. One possibility would be the `maximal temporal gauge' which 
requires to fix all temporal links to the identity except in one layer, 
but applying the mean field approximation would lead to a mean field that 
is not constant under time translations; this would not only be 
cumbersome, but probably also a poor approximation since it is violating 
a basic symmetry of the problem. We take instead the next simplest choice: 
we fix the temporal gauge field to be constant (`constant temporal' or 
`Polyakov gauge'). While the maximal temporal gauge does not lead to a 
nontrivial Faddeev-Popov determinant, going from that to the constant 
temporal gauge involves a nontrivial Jacobian (see appendix).

A problem that was noted already in the eighties concerns the temperature 
dependence of the `deconfining' phase transition. This is not represented 
appropriately by the leading mean field approximation if one uses an 
isotropic lattice and varies $T$ be varying $N_\tau$. We therefore fix 
(somewhat arbitrarily) $\beta$ and $N_\tau$ and introduce the temperature 
through anisotropy between spatial and temporal parameters, see 
Eqs.(\ref{e.YMa}),(\ref{e.det}). There we introduced two anisotropy 
parameters $\gamma_G$ and 
$\gamma_F$; in principle they should both be determined as a function of  
the single parameter $\gamma_{phys}$ by requiring space-time symmetry at 
$C=0$ and $T=0$. To leading order, however, we may set 
$\gamma_G=\gamma_F=\gamma_{phys}\equiv\gamma$; this is what was done in 
the computations in the appendix, since at this stage we cannot determine 
$\gamma_{phys}$ and the mean field computations are only meant to give a 
tentative picture of the phase structure.

The temperature is then related to $\gamma$ by
\be
aT=\frac{\gamma}{N_\tau}\,,
\end{equation}
where the lattice spacing $a$ is in principle determined by $\beta$. 
(Notice that there is now a nonzero minimal temperature.)

\begin{figure}[t]
\includegraphics[width=.9\columnwidth]{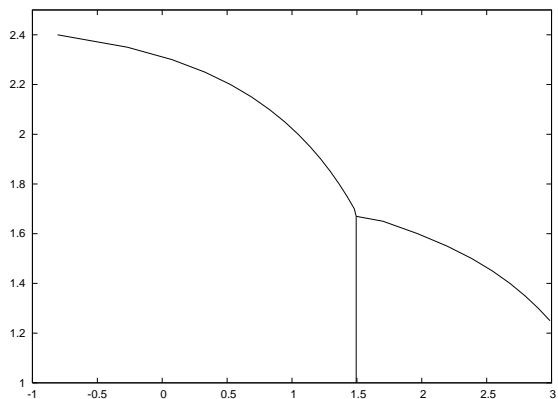}  
\caption{Mean field phase diagram (abscissa $\mu$, ordinate 
$\gamma=N_\tau\, a\,T$).}
\label{mf} 
\end{figure}

The mean field approximation is expressed in terms of two different mean 
fields $u$ and $v$ for the spatial and temporal gauge field links, 
respectively. In Fig.\ref{mf} we give an illustrative example, taken with 
$\beta=4$ and $N_\tau=6$. It shows a large `confinement' region for small 
$T$ and $\mu$ corresponding to the trivial fixed point mentioned above 
with both mean fields $u$ and $v$ vanishing. For larger $T$ or $\mu$ one 
crosses into a deconfined regime with both mean fields $u,v> 0$. In the 
lower right corner there appears in addition an intermediate phase with 
$u=0,\ v>0$. The field $v$ is close to its maximal value 1 wherever it 
is nonzero, whereas $u$ has smaller, varying values, depending on the 
region.

Of course the fact that the mean fields $u$ and $v$ are exactly zero in 
some regions is an artifact of the mean field approximation; according to 
earlier experience already the next approximation in the saddle point 
expansion would eliminate this feature. But qualitatively the mean 
field results indicate three phases in which different amounts of disorder 
are present: in the confined phase all the gauge fields are 
very much disordered, in the intermediate phase the Polyakov loops 
become ordered, while the spatial gauge fields remain disordered; 
finally there is the deconfined phase in which all the gauge fields show a 
high degree of order, but the Polyakov loops represented by $v$ more so than 
the spatial gauge fields represented by $u$. In the mean field picture we 
present here, increasing $\mu$ at fixed temperature, one first goes from 
the confined to the intermediate phase and then from there to the 
deconfined phase. This may be an artifact of the approximation and in 
reality the boundary between the intermediate and deconfined phases
may go upward. In any case, the simulations to be shown in the next 
section suggest that by making the chemical potential very large at fixed 
temperature we end up in the `half-ordered' phase. 

\section{Simulations and Results}
\label{numeric}

\subsection{Phase diagram}

As stated in the introduction, the model we are studying arises from the 
double limit $\kappa\to 0$ and $\mu\to 0$ of QCD, keeping $\zeta=\kappa 
\exp(\mu)$ fixed. It can be seen either as a laboratory to study QCD at 
large mass density near the quenched limit with a non-zero baryon density 
or as a model interesting by itself at any value of $\mu$ and $\kappa^2$, 
describing a dense system of heavy baryons.

The model still has a the sign problem that is getting more serious with 
increasing $\mu$. But for not too large values of $\mu$ and not too large 
lattices a local algorithm with a reweighting still converges in 
reasonable computer time, as will be shown explicitly below. Thus we are 
able to carry out simulations across large $\mu$ ``transitions'' at $T$ 
significantly below the deconfining temperature $T_c$ at $\mu=0$.

\begin{figure}[t] 
\includegraphics[width=.95\columnwidth]{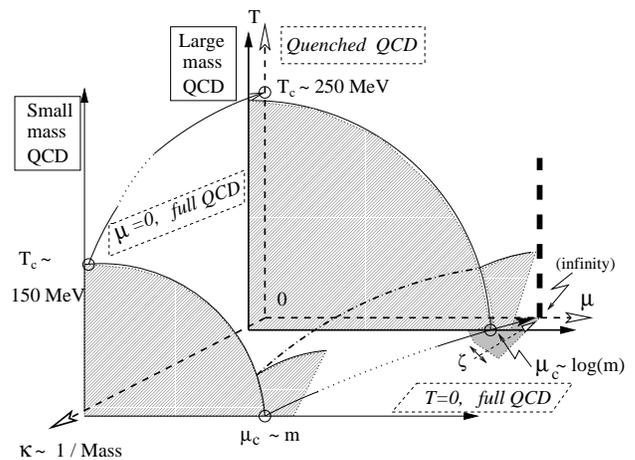} 
\caption{Tentative phase diagram in  $T$ and $\mu$ for various $\kappa$.} 
\label{ph_dia_8} \end{figure}

The tentative phase diagrams $T$ vs.$\mu $ are shown in Fig. \ref{ph_dia_8}. 
Here we show three planes: One corresponds to ``quenched'' QCD 
with a finite density of infinitely heavy quarks at $\kappa=0$.
This case has been studied for small $N_\tau$ in \cite{fktre,bky}.
At zero density we should find the first order phase transition of pure 
SU(3) Yang-Mills theory at $T_c\approx 250$ MeV. 

The plane in front is the region of $\kappa$ near the critical value
corresponding to masses that are small in lattice units. Here it has been 
found that there is only a crossover between confined and deconfined 
phases for all values of $\mu<\mu_c$, $\mu_c\approx 400$ MeV.
For $\mu \ge \mu_c$ one expects a sharp transition, curving down 
towards $T=0$ with increasing $\mu$ \cite{karrev}. 
It has been conjectured that at small 
$T$ above some value of $\mu$ a new phase exists, different from the 
deconfined (quark-gluon plasma) phase; this phase might be describable as 
a color superconductor and if the number of flavors is
$N_c=3$ ``color flavor locking'' (CFL) is expected \cite{arw}.  

Our model corresponds to a plane in between, i.e. small but 
positive $\kappa$, to be chosen below; as described in Section II, it is 
based on an expansion of the hopping parameter up to order $\kappa^2$. 
Since $\kappa$ is essentially proportional to $1/M$, our model contains some 
unquenched dymanics due to the fact that we are near but not in the 
quenched limit $\kappa=0$. We expect the phase diagram to be similar to 
the one for small mass just described. To check this is one of the 
purposes of this study.

\begin{figure}[t] 
\includegraphics[width=.95\columnwidth]{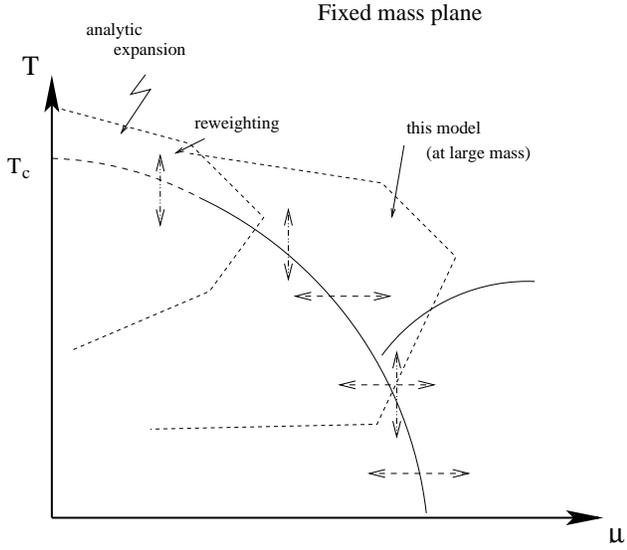}
\caption{Fixed 
mass plane phase diagram; dotted arrows indicate sequences of runs.} 
\label{ph_dia_9} \end{figure}

We are studying here for $\kappa=0.12$, mostly 
the region of high $\mu$, see Fig. \ref{ph_dia_9}. 
In this region
the phase diagram in temperature and chemical potential is expected to 
have a line 
of deconfinement transitions running into a
triple point at some nonzero $\mu$ and $T$. As 
mentioned above, at this point two further phase transition lines branch 
off, separating the new ``color superconducting'' or color-flavor locked
phase from the quark-gluon plasma as well as the confined hadronic phase.
It has been a long standing challenge for lattice QCD to explore this region.

\begin{figure}[t]
\includegraphics[width=.9\columnwidth]{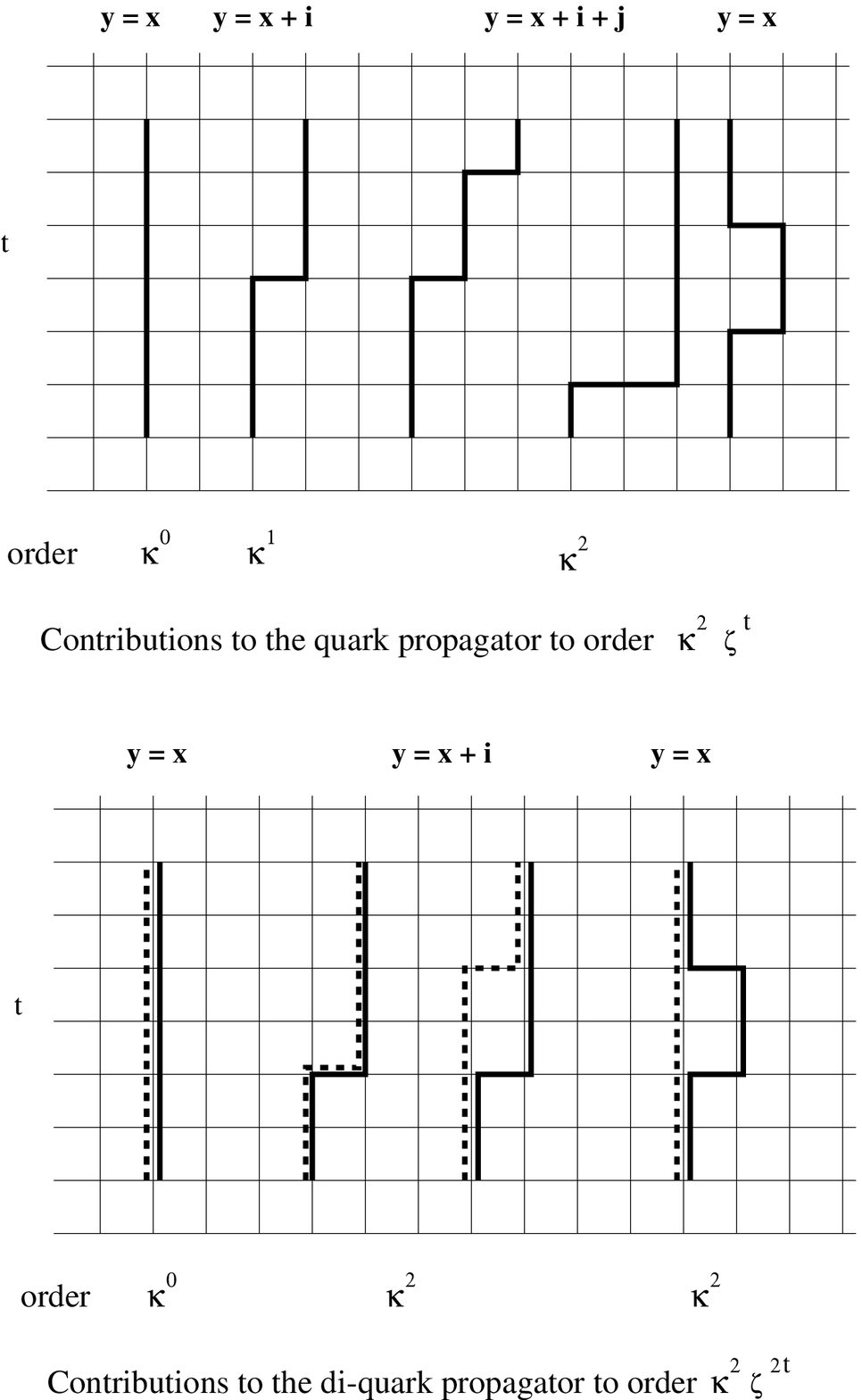}  
\caption{Paths contributing to quark and diquark ``propagators''.} 
\label{prop_2} 
\end{figure}

\subsection{Observables}
We measure several observables under the variation of $\mu$ and 
$T$, to check the properties of the different phases for small $T$ and large
$\mu$. In the following we specialize to $N_c=3$. The observables are: 
the Polyakov loop,
\bea
\langle P \rangle =  \langle \frac{1}{3\,N_\sigma^3} \sum_{\vec x} \Tr {\cal 
P}_{\vec x}\rangle = \langle \frac{1}{N_\sigma^3} \sum_{\vec x} P_{\vec x}\rangle  
\,,
\eea
and its susceptibility
\bea
\chi_P=\sum_{\vec y} \left( \langle P_{\vec x}\, P_{\vec y}\rangle
- \langle P_{\vec x}\rangle \langle P_{\vec y}\rangle\right)\,,
\eea
the (dimensionless) baryon number density $n_B$, 
\bea
n_B= \sum_f \frac{n_{b,f}}{T^3} \, ,
\eea
where the contribution of each flavor is:
\bea
&& \frac{n_b}{T^3} = \frac{N_\tau^3}{3 N_\sigma^3} \hat{n} \, , \qquad
\hat{n} = \hat{n}_0 + \hat{n}_1 \, , \nonumber \\
&& \hat{n}_0 = \langle \frac{\partial}{\partial \mu} {\cal Z}_F^{[0]} 
\rangle 
\approx 2 C \langle \sum_{\vec x} \Tr {\cal P}_{\vec x} \rangle \nonumber \\
&& \hat{n}_1 =\langle \frac{\partial}{\partial \mu} 
\bigg( \frac{{\cal Z}_F^{[2]}}{ {\cal Z}_F^{[0]}} \bigg)  \rangle 
\approx 2 C \kappa^2 \langle \sum_{\vec x} \Tr {\cal P}_{{\vec x},i,t,t'} 
\rangle \, ,
\eea
with the corresponding susceptibility
\bea 
\chi_{n_B} = \langle n_B^2 \rangle - \langle n_B \rangle^2\, ,
\eea
the spatial and temporal plaquettes $\frac{1}{3}\Tr\,P_{\sigma \sigma}$, 
$\frac{1}{3}\Tr\,P_{\sigma \tau}$ 
and the topological susceptibility $\chi_{top}= \langle
Q^2_{top}\rangle/(N_\sigma^3 N_\tau)$. The topological
charge was measured using  an
improved field theoretical
formula based on five Wilson loops \cite{mnp}. In order to check the 
character of the conjectured third phase we also measure the diquark - diquark
correlators
\bea
&& C_{(qq)}(\tau) = (\delta_i^a \delta_j^b + \xi \delta_j^a \delta_i^b )
(\delta_k^c \delta_l^d + \xi \delta_l^c \delta_k^d ) \nonumber \\
&& \times \sum_{x,y,t} \langle[\psi_i^a {\cal C} \psi_j^b(x,t)] 
[\psi_l^c {\cal C} \psi_k^d(y,t + \tau)]^\star \rangle \nonumber \\
&& = (\delta_i^a \delta_j^b + \xi \delta_j^a \delta_i^b )
(\delta_k^c \delta_l^d + \xi \delta_l^c \delta_k^d ) \nonumber \\
&& \times \sum_{x,y,t} \bigg\{ W^{-1}_{ik;ac}(x,t;y,t+\tau) {\cal C}^T 
W^{-1,T}_{jl;bd}(x,t;y,t+\tau) {\cal C}  \nonumber \\
&& - W^{-1}_{il;ad}(x,t;y,t+\tau) {\cal C}^T 
W^{-1,T}_{jk;bc}(x,t;y,t+\tau) {\cal C} \bigg\} \, ,
\label{dq}
\eea
where $W^{-1}$ is the quark propagator measured in maximal temporal gauge, 
${\cal C}$ the charge conjugation 
matrix $\{a, \cdots; i, \cdots \}$ the color the flavor indices, respectively,
and we have dropped the (summed over) Dirac indices. $\xi$ is a parameter 
allowing various combinations of color-flavor ``locking'' (see \cite{arw}).
Fig. \ref{prop_2} shows the contributions
to order $\kappa^2 \xi^{2 t}$ to quark and di-quark propagators. The
corresponding susceptibility is the integral of $C_{qq}$.

\subsection{Algorithm and simulations}
We use the Wilson action and Wilson fermions within a reweighting 
procedure. The updating is performed with a local Boltzmann factor which 
only leads to a redefinition of the ``rest plaquette'':
\begin{multline}
B_0(\{U\})\equiv\prod_{Plaq} e^{\frac{\beta}{3} Re \Tr Plaq} \times \\
\times \prod_{\vec x}  \exp\bigg\{ 
2 C Re \Tr \bigg[ {\cal P}_{\vec x} + \kappa^2 \sum_{i,t,t'}
{\cal P}^{0,1}_{\vec x,i,t,t'} \bigg] \bigg\} \, .
\end{multline}
The weight (global, vectorizable) is
\begin{multline}
w(\{U\})\equiv \prod_{\vec x} \exp\bigg\{ 
\!\!  -2 \, C\, Re \Tr \bigg[ {\cal P}_{\vec x} + \kappa^2 
\sum_{i,t,t'}
P^{0,1}_{\vec x,i,t,t'} \bigg] \bigg\}  \\ 
\times {\cal Z}^{[2]}_F(\{ U \}) \, ,
\end{multline}
such that, 
\begin{equation*}
w\,B_0\, = \, B\,\equiv \prod_{Plaq} e^{\frac{\beta}{3}Re \Tr 
Plaq}\,{\cal Z}^{[2]}_F(\{ U \}) \, .
\end{equation*} 
Averages are calculated by reweighting according to 
Eqs.(\ref{e.b0w0}), (\ref{expw}). 

We have employed the Cabibbo-Marinari 
heat-bath procedure mixed with over-relaxation. This updating already 
takes into account part of the $\mu > 0$ effects and the generated 
ensemble can thus have a better overlap with the true one than an updating 
at $\mu = 0$. One can also use an improved $B_0$, to be taken care of by a 
supplementary Metropolis check. Anisotropy can be straightforwardly 
introduced. 
Notice that extracting a factor like $B_0$ may also improve 
convergence of 
full QCD simulations at $\mu > 0$. 

The simulations are mainly done on lattice $6^4$ for $n_f=1,3$ degenerate 
flavors (any mixture of flavors can be implemented). The $\kappa$ 
dependence has been analyzed in \cite{hdm01}. Here we set $\kappa=0.12$ 
(rather ``small'' bare mass $M_0 = 0.167$) which drives the $1/M^2$ 
effects in the baryonic density to about $50 \%
$. The task we have set to 
ourselves is primarily to explore the phase structure of the model at 
large chemical potential and ``small'' temperature and we
accordingly vary $\mu$
 and $\beta$. We also 
want to check the behavior of bulk properties around the prospective 
``transition'' line.

\subsection{Results and discussion}

\begin{figure}
\includegraphics[width=\columnwidth]{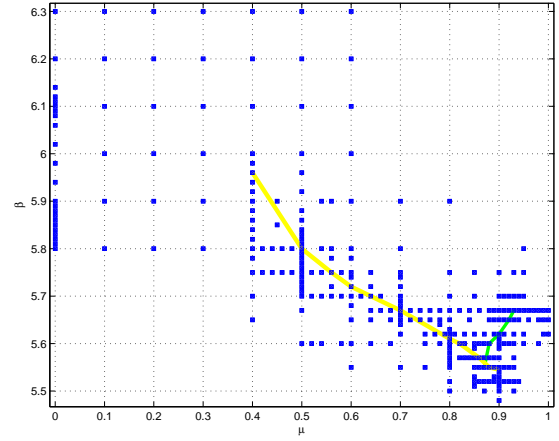}  
\caption{Data taken in the plane $\beta$ vs. $\mu$ for fixed $\kappa=0.12$. } 
\label{data} 
\end{figure}

The algorithm works reasonably well over a large range of parameters 
 even at small temperature.
The model permits to vary $\mu$, $\kappa$, $\beta$ as independent parameters 
and it is reasonably cheap to measure various correlations. The region we 
have analyzed on a $6^4$ lattice with $n_f=3$ is shown in Fig. \ref{data}. 
We have also run simulations on larger and smaller lattices, but we 
decided to base our discussion on the $6^4$ data and also on one value 
$\kappa=0.12$. For $8^3\times 4$ and $8^4$ lattices the $n_f=3$ data are
not good enough in the (interesting) high $\mu$ region and therefore
we do not introduce them in the discussion.
All results are expressed in lattice units, and we simulate the temperature
variation by varying $\beta$ according to (\ref{e.temp0})
 with $\gamma_{phys}=1$. 
To avoid the problem of fixing the scale
we shall consider $T/T_c$ with $T_c$ of the $\mu=0$, pure gauge theory.  
We shall comment on all this in the conclusions.

\begin{figure*}
\includegraphics[width=0.85\textwidth]{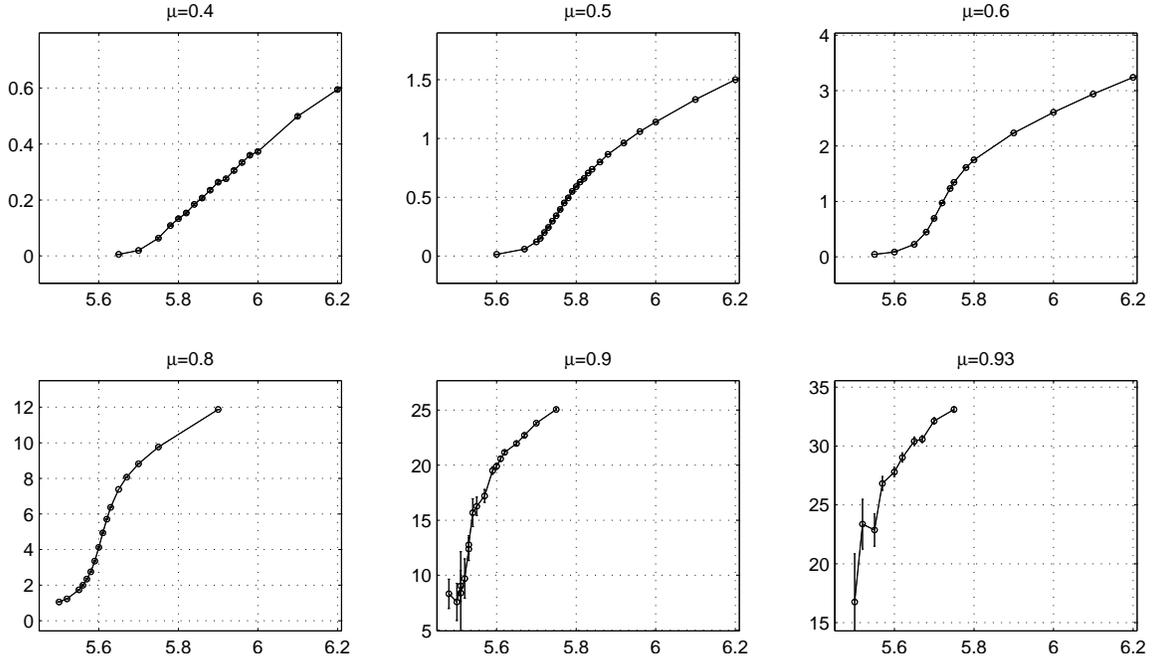}  
\caption{Baryonic density vs. $\beta$ at fixed $\mu$.} 
\label{bareps} 
\end{figure*}
\begin{figure*}
\includegraphics[width=0.85\textwidth]{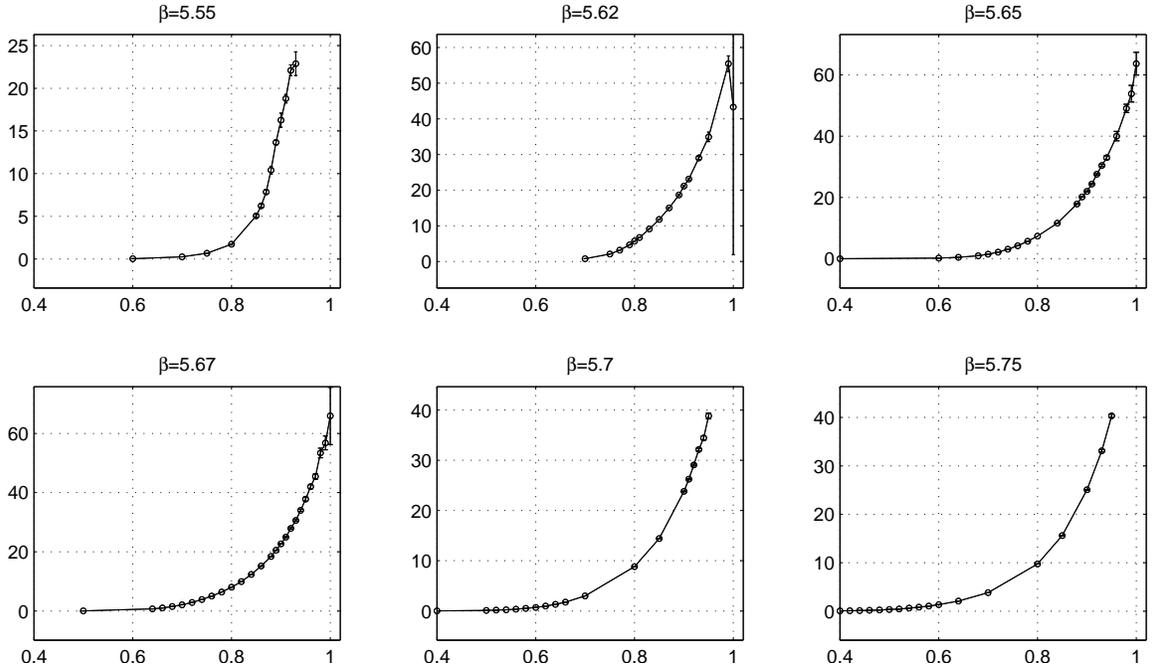}  
\caption{Baryonic density vs. $\mu$ at fixed $\beta$.} 
\label{bareps2} 
\end{figure*}
\begin{figure}
\includegraphics[width=\columnwidth]{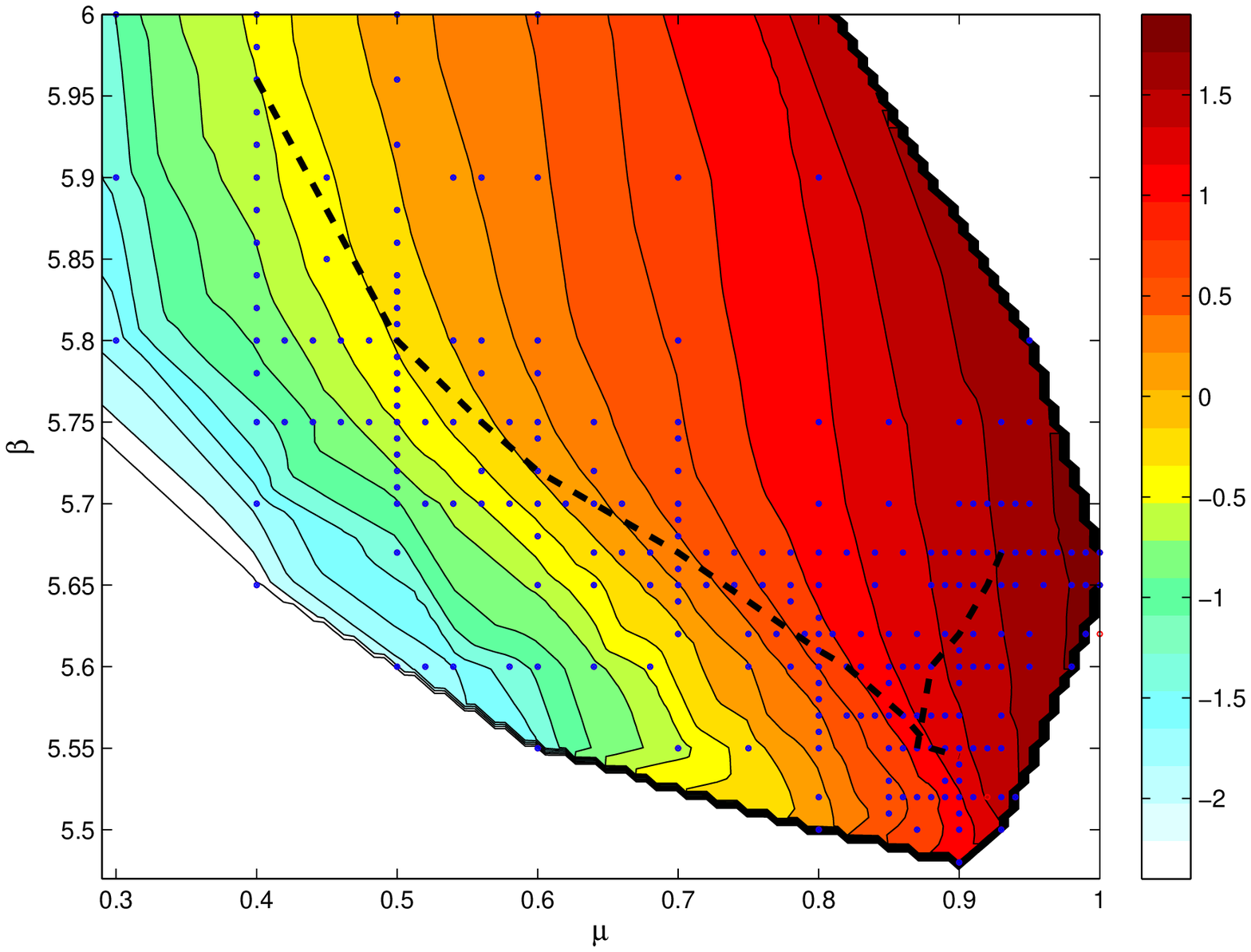}  
\caption{Landscape of the baryonic density. 
The color scale (right) is based on $\log_{10} (n_B)$.} 
\label{densR} 
\end{figure}

In Fig. \ref{bareps} we show the behavior of the baryonic density $n_B$ with 
$\beta$ at fixed $\mu$ values.
 We see 
at the different values of $\mu$ inflection points 
(maximal slope) in $\beta$ indicating 
possible qualitative changes of behavior suggesting transitions from low 
to high temperature phases. In Fig. \ref{bareps2}  we vary $\mu$ at 
several fixed $\beta$ values and see the expected rapid increase of 
$n_B$ with $\mu$, indicating that we do not see yet saturation effects
\cite{hands}. Finally, in Fig. \ref{densR} we show the 
``landscape'' of the real part of the baryon density (while the imaginary 
part is compatible with zero inside the statistical errors, as it should be). 

\begin{figure}
\includegraphics[width=\columnwidth]{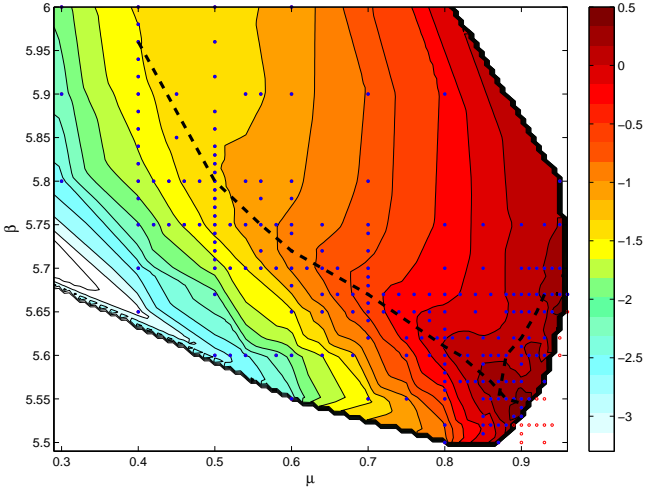}  
\caption{Landscape of the baryon density susceptibility.
The color scale (right) is based on $\log_{10} (\chi_{n_B})$.} 
\label{chemland} 
\end{figure}

A clearer view of the situation is provided by looking at the 
``landscape'' of the susceptibility of the baryon density, which is shown 
in Fig. \ref{chemland}. A ridge is clearly visible, 
highlighted by a dashed black line. A second line (dotted) will be 
explained later.

\begin{figure*}
\includegraphics[width=0.85\textwidth]{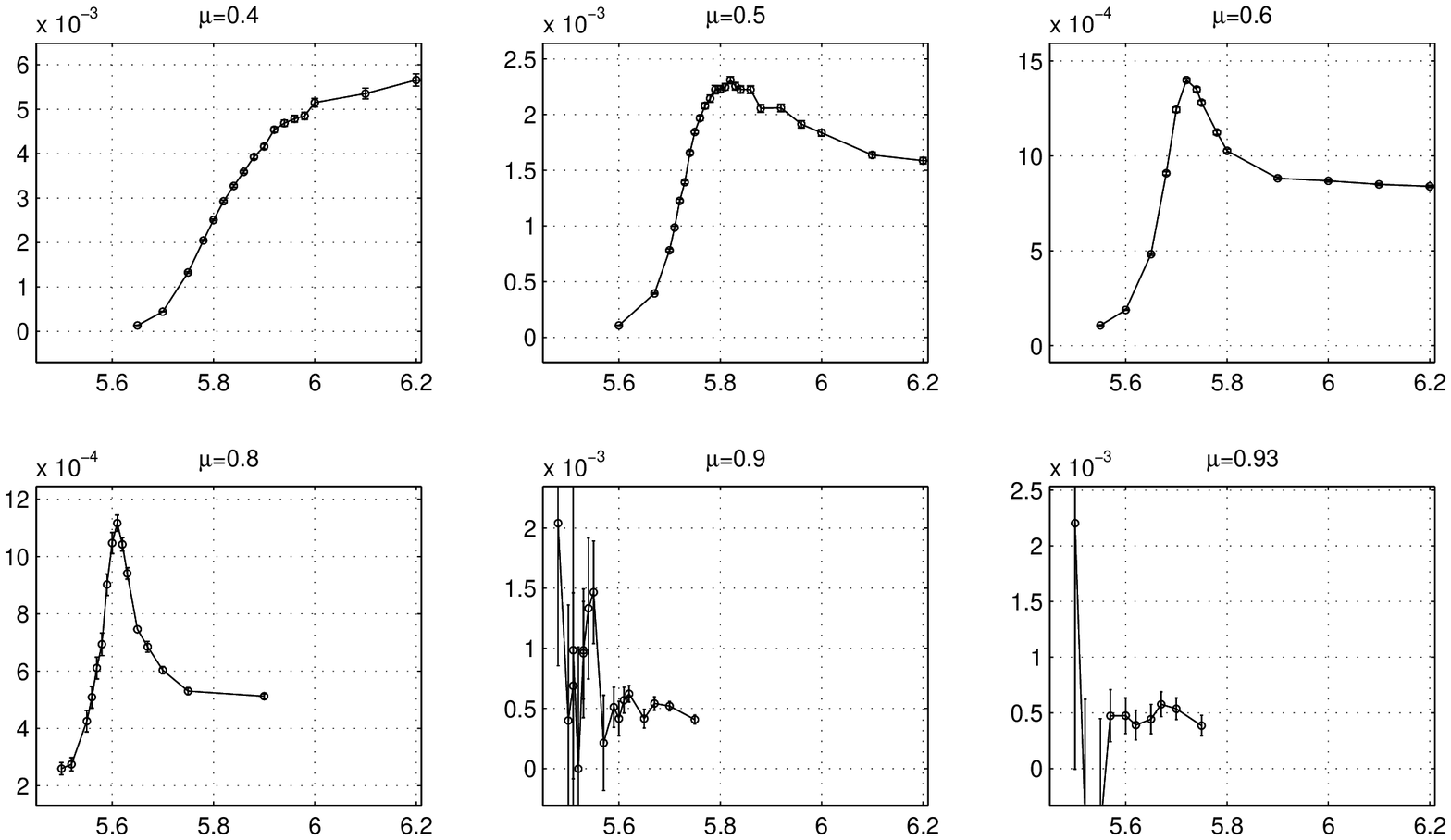}  
\caption{Polyakov loop susceptibility vs. $\beta$ at fixed $\mu$. } 
\label{poleps} 
\end{figure*}

\begin{figure*}
\includegraphics[width=0.85\textwidth]{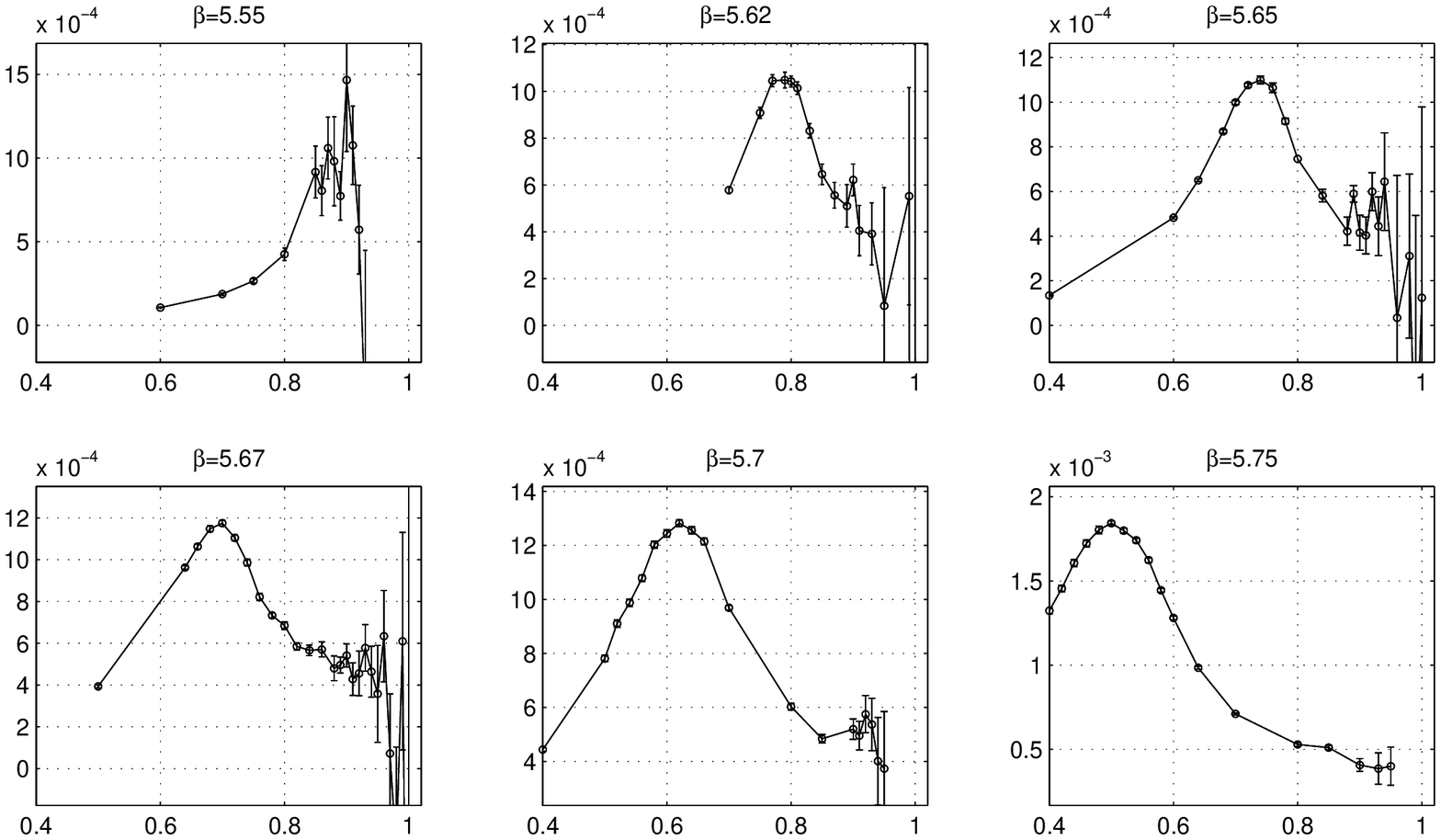}  
\caption{Polyakov loop susceptibility vs. $\mu$ at fixed $\beta$. } 
\label{poleps2} 
\end{figure*}

\begin{figure} 
\includegraphics[width=\columnwidth]{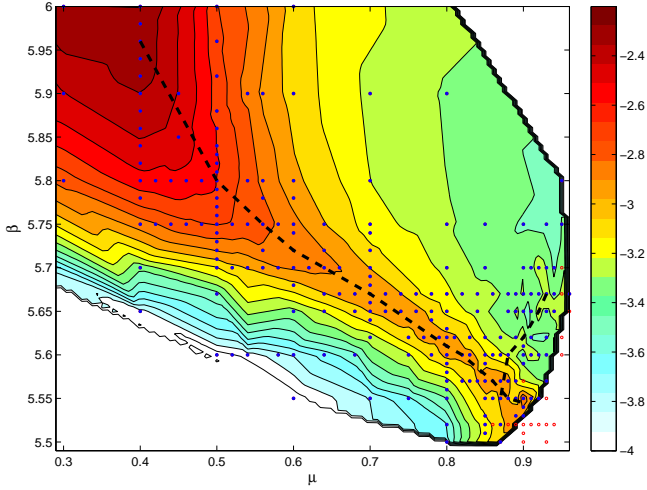}  
\caption{Landscape of the Polyakov loop susceptibility.
The color scale (left) is based on $\log_{10} (\chi_P)$} 
\label{polland} 
\end{figure}

\begin{figure} 
\includegraphics[width=\columnwidth]{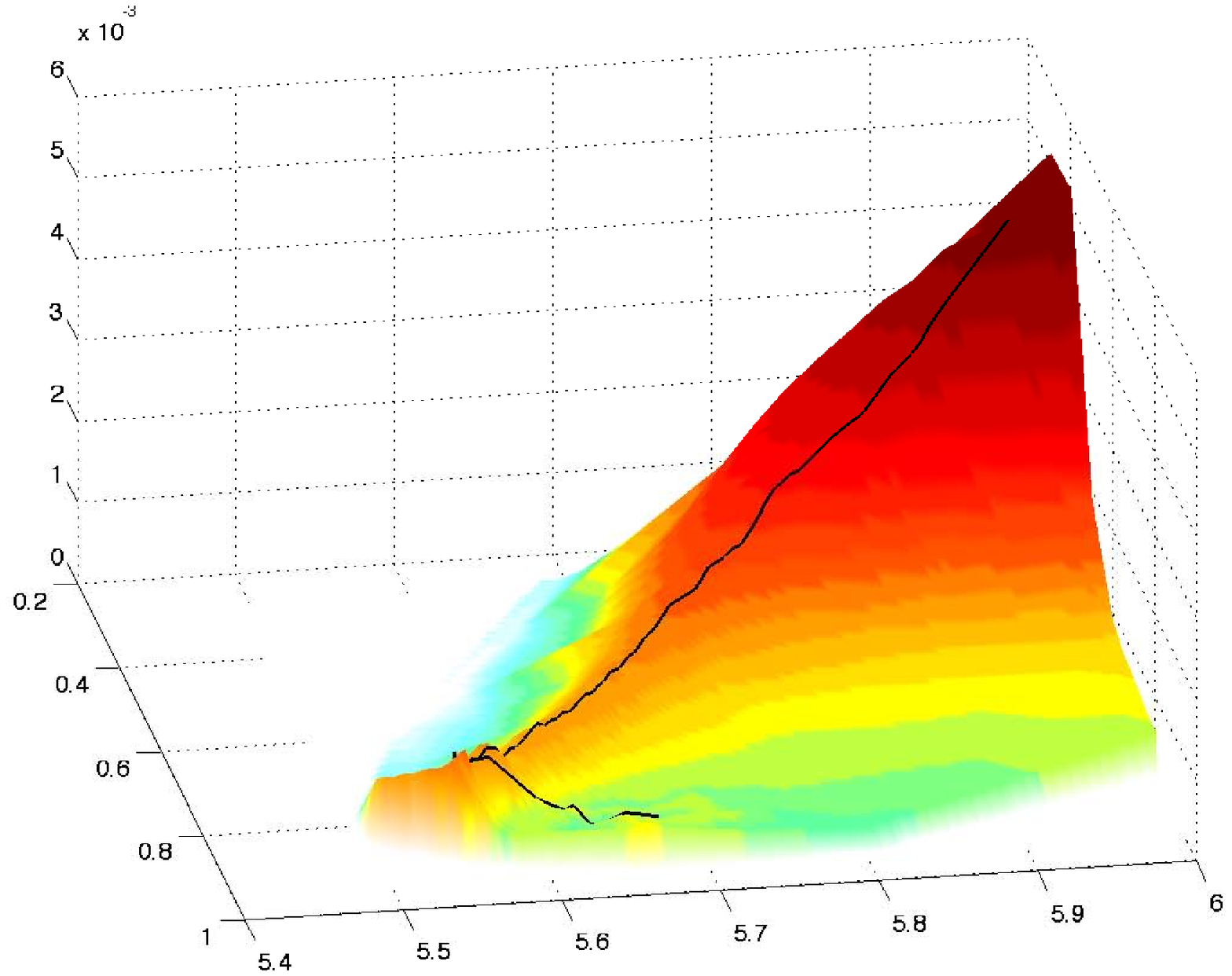}  
\caption{$3d$ view of Fig. \ref{polland}.} 
\label{3dpolland} 
\end{figure}

The main variation in the baryon density is an exponential growth with 
$\mu$. This masks to a certain extent the finer structure. We found it 
therefore advantageous to look at the Polyakov loops and their 
susceptibility. In Fig. \ref{poleps} we show this susceptibility at fixed 
$\mu$ vs $\beta$ and in Fig. \ref{poleps2} at fixed $\beta$ vs. $\mu$, and 
in Figs. \ref{polland} and \ref{3dpolland} the corresponding landscape. 


The plots of the Polyakov susceptibility show quite clearly maxima 
indicating possible transitions or crossovers. In the landscape 
Figs. \ref{polland} and \ref{3dpolland} one of 
these maxima shows up as a well defined ridge, indicated by a dashed black 
line. It shows only a moderate slope in $\mu$, which explains why the 
maxima are more pronounced when we vary $\beta$ at fixed $\mu$ than vice 
versa. The broadening of this ridge at small $\mu$ as well as of the 
maximum in Fig. \ref{poleps} is responsible for the loss of a sharp 
transition signal at small $\mu$. These figures clearly show that the
transition at fixed $\mu = 0.50$ is less steep than the one at
$\mu=0.80$. Presumably at $\mu < \sim 0.6$ we are dealing with a crossover,
whereas at large $\mu$ the signal is more compatible with
 a real phase transition. Notice that
changing $\beta$ at fixed $\mu$, we cross the transition line at a more
oblique angle at smaller $\mu$, but the broadening of the ridge and
loss of a transition signal is a genuine effect, as can be seen from 
Figs. \ref{polland} and \ref{3dpolland}.

A second ridge branching off from this main ridge at large $\mu$, 
highlighted by a dotted line is suggested by looking at the level lines in 
Fig. \ref{polland} and corresponds to the second maximum suggested at 
large $\mu$ in Fig. \ref{poleps2}. This may indicate the appearance of the new 
phase at large $\mu$ and small $T/T_c$ discussed above.
\begin{figure*} 
\includegraphics[width=0.8\textwidth]{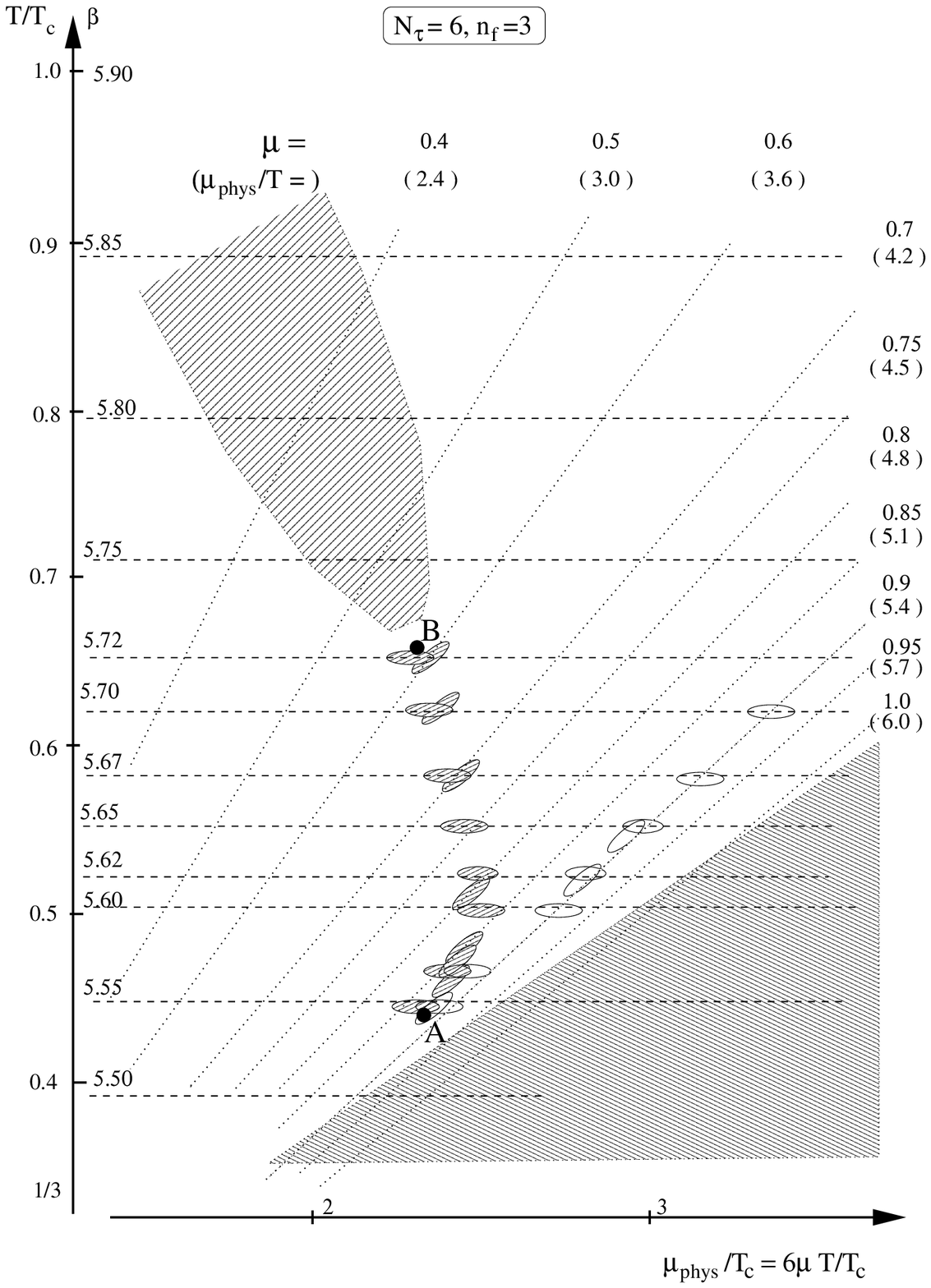} ~~~~ 
\caption{Phase diagram in the  $\beta$ (or $T/T_c$) - $\mu_{phys}/T_c$ QCD
plane. The dotted straight lines correspond to constant $\mu$, 
the dashed ones to 
constant $\beta$. The blobs, shadowing and other features 
are explained in the text.} 
\label{phtm4} 
\end{figure*} 

We use the results for the Polyakov loop susceptibility to estimate the 
possible position of the transition points in the $\beta$ vs $\mu$ plane;
to go half way toward a possible physical interpretation 
the positions determined in this way are indicated by the blobs in the
diagram $T/T_c$ vs. $\mu_{phys}/T_c$ of Fig. \ref{phtm4}, 
where  
$\mu_{phys}= \mu / a(\beta) = N_{\tau} \mu T$ 
and the relation between $\beta$ and $T/T_c$
has been roughly estimated from the $\mu=0$ quenched QCD with
$N_\tau=6$ (we shall comment on this point in the conclusion section).
In this figure the axis of the blobs
indicate the search lines in the simulation. The shaded blobs correspond
to the rather unambiguous `deconfining' signal observed for 
$\mu > \sim 0.6$ ($\beta <\sim 5.72$). The `transition' line suggested 
by this signal
 starts at the lower point A on the figure, located at  
$\beta\simeq5.55,\,\mu\simeq 0.88$, i.e., with our rough estimation
$\mu_{phys}/T_c \simeq 2.4,\,T/T_c\simeq 0.45$ (below which we could no
longer obtain reliable data) and ends at the point B located near
 $\beta\simeq 5.72,\,\mu\simeq 0.6$, i.e., with our rough estimation
$\mu_{phys}/T_c \simeq 2.3,\,T/T_c\simeq 0.65$. Above this point the 
signal becomes ambiguous.
But one should keep in mind that moving along lines of fixed $\mu$ across 
a broad ridge, the maximum in general is shifted with respect to the 
ridge (in our case to lower $\beta$ values), the location of a transition 
becomes somewhat blurred, in accordance with the claim that here we are 
dealing with a crossover and not a phase transition.  In Fig. \ref{phtm4} 
we shaded the upper, `broad ridge region' above B where the maximum at 
fixed $\mu$ or $\beta$ deviates significantly from the location of the 
ridge, which can be easily understood from the landscape Fig. \ref{polland}. 
Notice that since we keep $\kappa$ fixed $\mu=0$ does not represent
the pure Yang Mills theory therefore we did not try to go to this limit.
The white blobs correspond to the more volatile, possible 'transition' 
branching off near point A at large $\mu$, whose signal is strongly 
affected by fluctuations. 
We also shaded the region at high $\mu$ in the lower right hand corner,
where we could not obtain 
reliable data due to the sign problem.  

The picture emerging from the data is thus the following: for $\mu < 0.5 - 
0.6$ ($\mu_{phys}/T\sim 3$)  there is only a broad crossover, while for 
$0.6 < \mu < 0.9$ ($3.6 < \mu_{phys}/T < 5.3$) there is evidence of a 
sharper crossover or transition at a value $\mu_c$ depending on $\beta$. 
Moreover, for $\mu \simeq 0.9$ there is some evidence of the presence of 
the second transition even though this evidence is much weaker than the 
other one because at larger values of $\mu$ the fermion determinant 
strongly oscillates and, indeed, the usual sign problem manifest its 
effects.

To get some further insight into the nature of the different regimes or 
phases we also wanted to look at the distribution of the values of the 
Polyakov loop in the complex plane. At first we considered the 
`histograms' corresponding to the following mathematical expression:
\begin{multline}
H_\Delta(x,y)= \\
\left\langle \Theta_{\Delta,x}  
\left(\frac{Re(w\,P_{\vec x})}{\langle w\rangle_0}\right)   
\,\Theta_{\Delta,y}\left(\frac{Im(w\,P_{\vec x})}{\langle 
w\rangle_0}\right)
\right\rangle_0
\label{eq:histogram}
\end{multline}
where $\vec x$ is any point in the spatial lattice and  
$\Theta_{\Delta,s}(t)$ is the function which is 1 if $|t-s|\le 
\Delta/2$ and 0 otherwise (the arguments $x,y$ in $H$ should not be
confounded with space-time points). For the figures we used $20\times 20$ 
bins choosing $\Delta$ accordingly.These quantities have the advantage 
that they are positive, because they use the expectation values $\langle 
.\rangle_0$ determined by the positive Boltzmann factor $B_0$ (see Eq. 
\ref{e.b0w0}); therefore they can be interpreted as probability 
distributions. But their disadvantage is that they depend on the choice of 
$B_0$. It should also be noted that they describe not really the 
distribution of the Polyakov loops themselves, but rather the product of 
the Polyakov loop with the weight factor $w$; for this reason absolute 
values larger than 1 are possible and actually occur, as we will see.

\begin{figure*}
\includegraphics{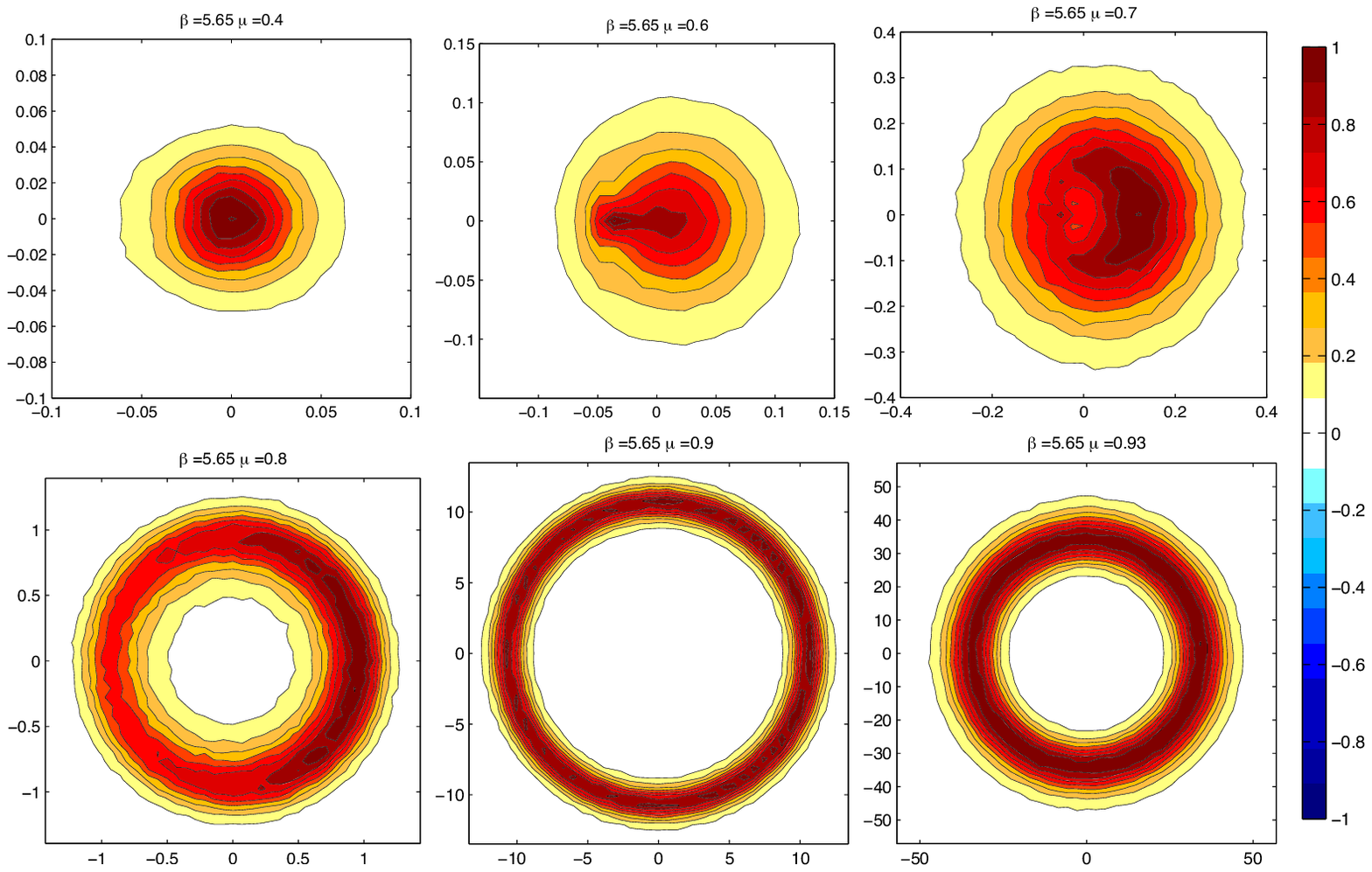}
\caption{Polyakov loop `histogram' $H_\Delta(x,y)$ of 
eq.\ (\ref{eq:histogram}) vs. $\mu$ at $\beta=5.65$.} 
\label{histobeta565} 
\end{figure*}

\begin{figure*}
\includegraphics{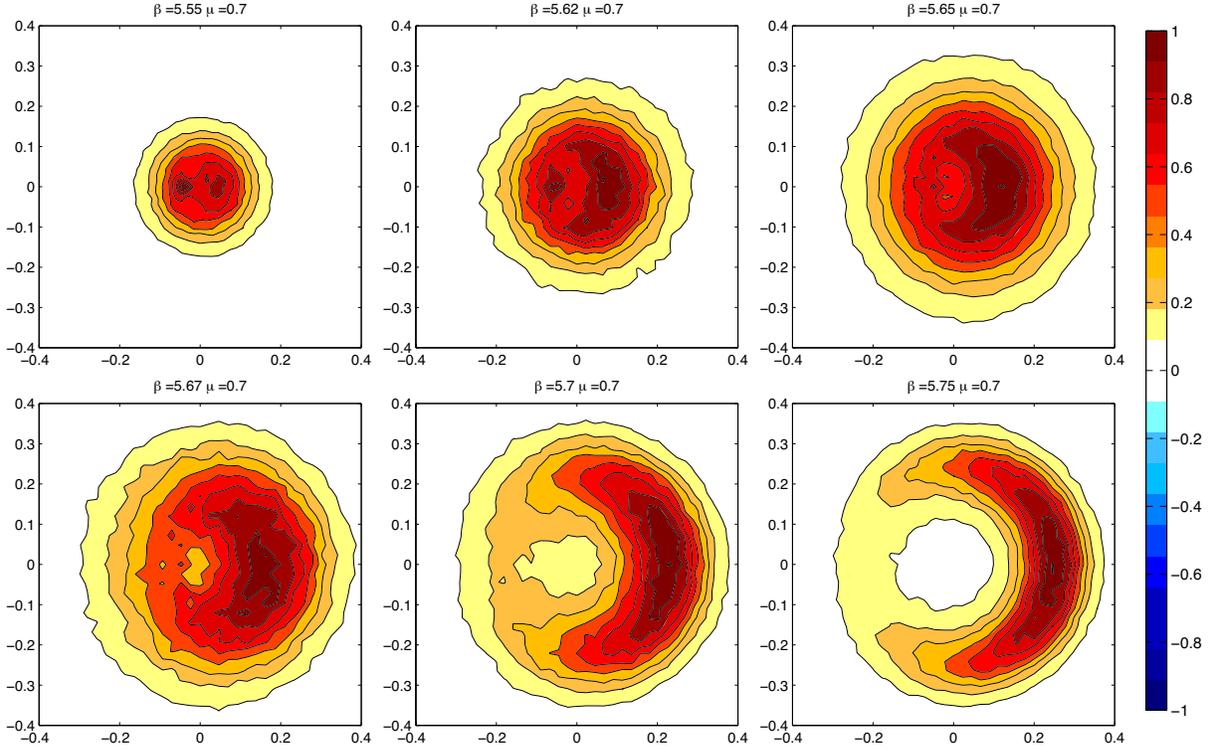}
\caption{Polyakov loop `histogram' $H_\Delta(x,y)$ of 
eq.\ (\ref{eq:histogram}) vs. $\beta$ at $\mu=0.70$.} 
\label{histomu070} 
\end{figure*}

As an example, see Fig. \ref{histobeta565} and Fig. \ref{histomu070} that 
represent the histogram of $H_\Delta$ at different 
values of $\mu$ at $\beta=5.65$ and different values of $\beta$ at 
$\mu=0.70$, respectively.  These figures show different behavior of this 
observable in accordance with the transition lines indicated in 
Fig. \ref{phtm4}. In fact in Fig. \ref{histobeta565} one can discern three 
different regions: the first one corresponds to $\mu < 0.6$, where the 
Polyakov loops are concentrated in a small region around zero with only a 
slight preference for positive real parts; in the second region, for $0.6 
< \mu < 0.9$ the Polyakov loops become considerably larger, favoring 
positive real parts in a significant way, while finally for $\mu > 0.9$ the 
Polyakov loops (times weight) becomes quite large, but are distributed 
almost symmetrically around the origin.

This picture can be corroborated by looking at Fig. \ref{histomu070}, 
which according to Fig. \ref{phtm4} should only show one transition. One 
can see a change of behavior around the point $\beta=5.65$ (which also 
occurs in Fig. \ref{histobeta565}): The Polyakov loops become somewhat 
larger with a distribution more heavily favoring positive real parts; we 
interpret this as the transition from a confined to a deconfined phase.


A `distribution' independent of the choice of $B_0$ can be defined by 
considering
\bea
T_{\Delta}(x,y)=\left\langle\Theta_{\Delta,x}(Re P_{\vec x})\,
\Theta_{\Delta,y}(Im P_{\vec x})\right\rangle\, ,
\label{eq:Tdistribution}
\eea
which means adding the weights of all configurations producing a
$P_{\vec x}$ value in a given bin
 $|Re P_{\vec x}-x|\leq \Delta/2\,,
\,|Re P_{\vec y}-y|\leq \Delta/2$.
Because now the ``expectation value'' $\langle .\rangle$ refers to the 
complex ``Boltzmann factor'' $B$ (see Eq. \ref{e.b0w0}), $T_\Delta$ is 
complex and does not represent a probability distribution. But for small 
$\Delta$ we have
\bea
\langle P \rangle\approx\sum_{x,y} (x+iy) T_\Delta(x,y)\,,
\label{approx}
\eea
where the sum runs over a lattice with lattice constant $\Delta$ in the 
$xy$-plane. Since the expectation value of $P$ is real, $Re T_\Delta$ has 
to be even and $Im T_\Delta$ odd in $y$. 

\begin{figure*}
\includegraphics{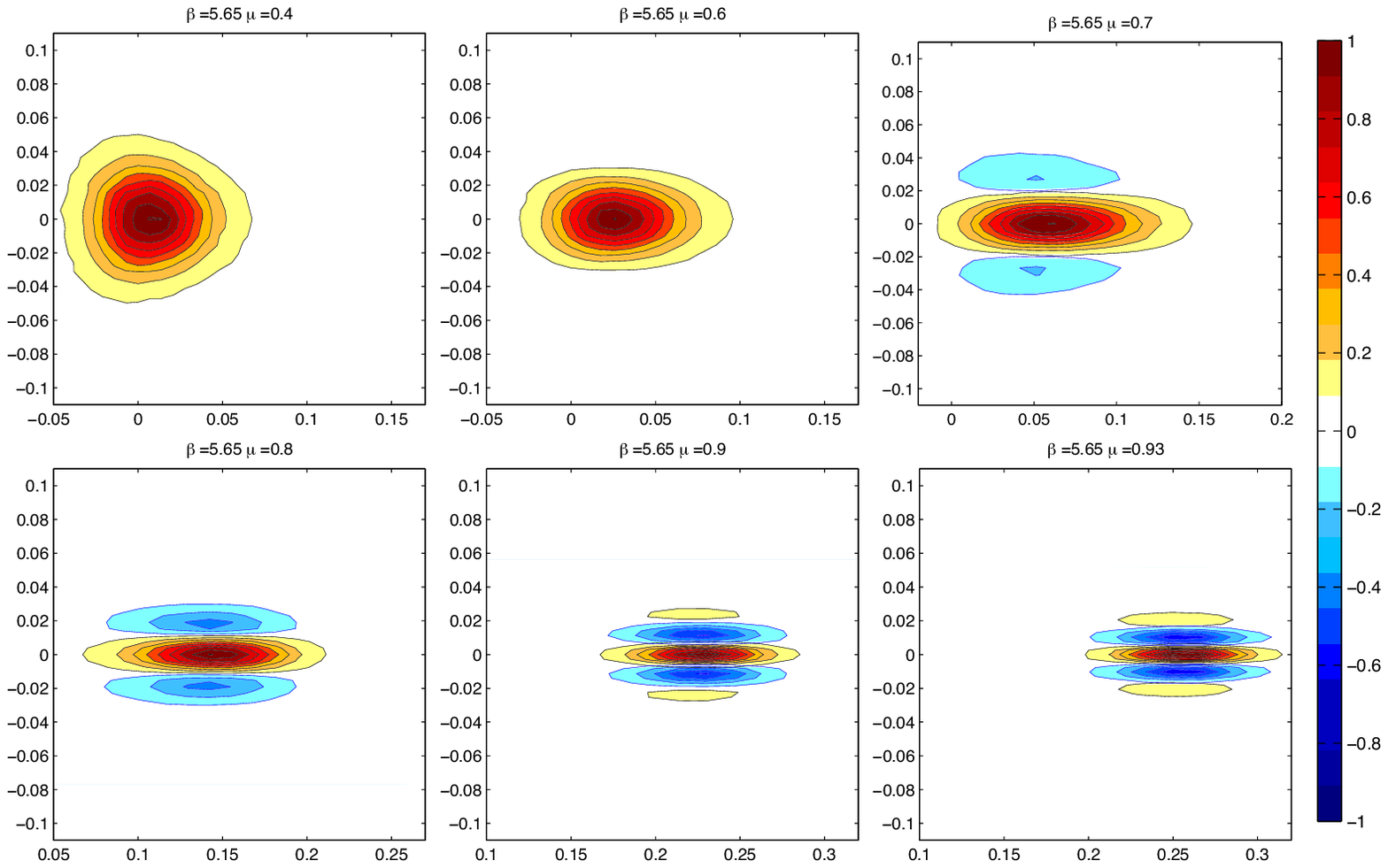}
\caption{Real part of the Polyakov loop `distribution'
$T_{\Delta}(x,y)$ of eq.\ (\ref{eq:Tdistribution}) 
vs. $\mu$ at $\beta=5.65$ fixed.} 
\label{rhismu} 
\end{figure*}

\begin{figure*}
\includegraphics{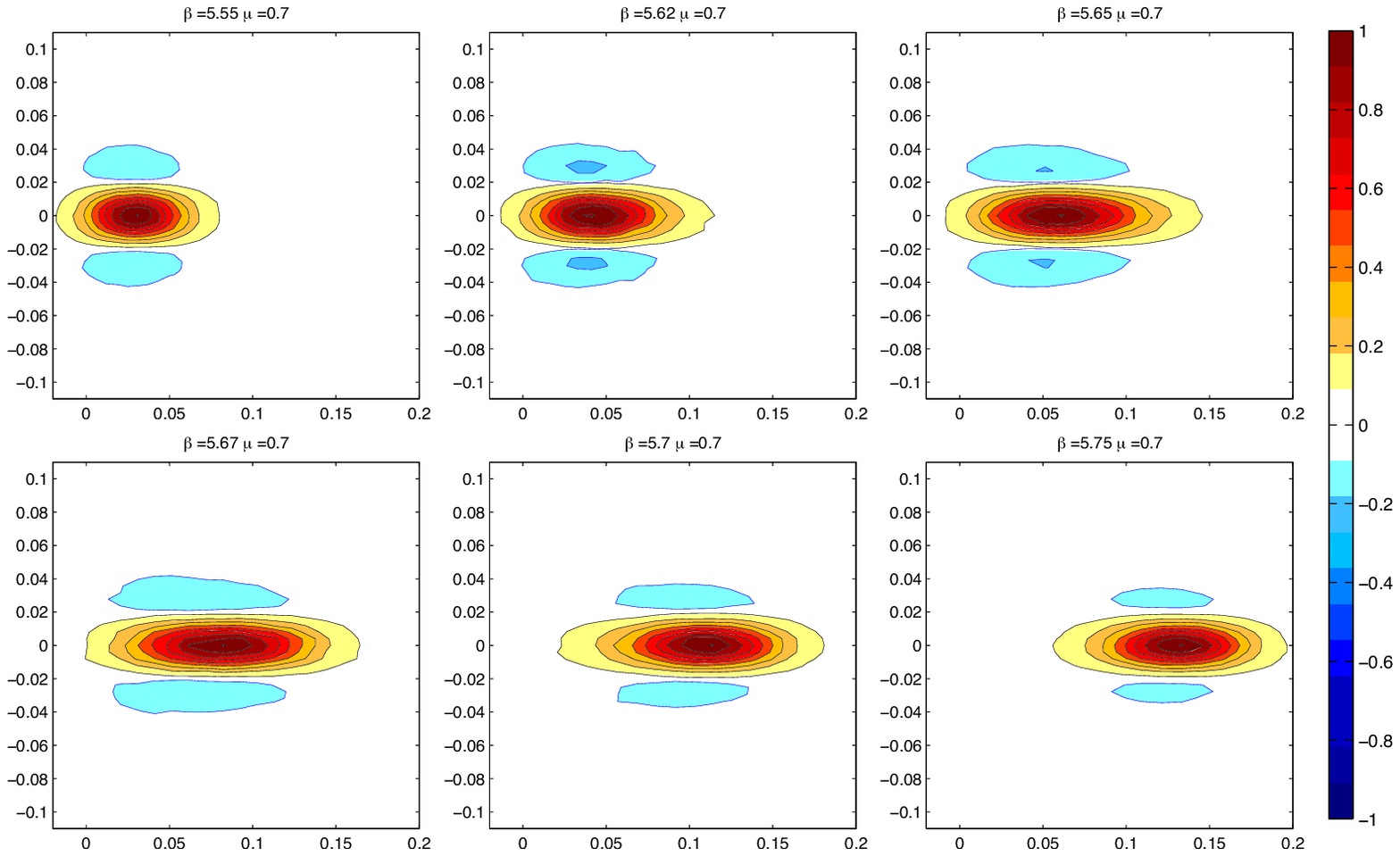}
\caption{Real part of the Polyakov loop `distribution'
$T_{\Delta}(x,y)$ of eq.\ (\ref{eq:Tdistribution})
vs. $\beta$ at $\mu=0.70$ fixed. } 
\label{rhisb} 
\end{figure*}

We give some representative figures showing the behavior of $T_\Delta$ 
across the putative transitions, for the same parameters as before. Fig. 
\ref{rhismu} shows $Re T_\Delta$ for $\beta=5.65$ for various increasing 
values of $\mu$. Again we should observe the crossing of two of the 
putative transition lines. The transition signals are not very strong, but 
we can observe that for $\mu<0.7$ negative real parts are present, which 
disappear for $\mu\ge 0.7$; at $\mu\ge 0.9$ the real parts become 
considerably larger again, reaching values of $0.3$. Fig. \ref{rhisb} 
shows $Re T$ at $\mu=0.7$ for increasing values of $\beta$. Here the 
parameters are such that we should observe only the transition between the 
hadronic and plasma phases. The indication for this is again that the real 
parts touch the origin for $\beta\le 5.65$, whereas for $\beta>5.65$ they 
increase to positive values, but staying below $0.2$.

\begin{figure*}
\includegraphics{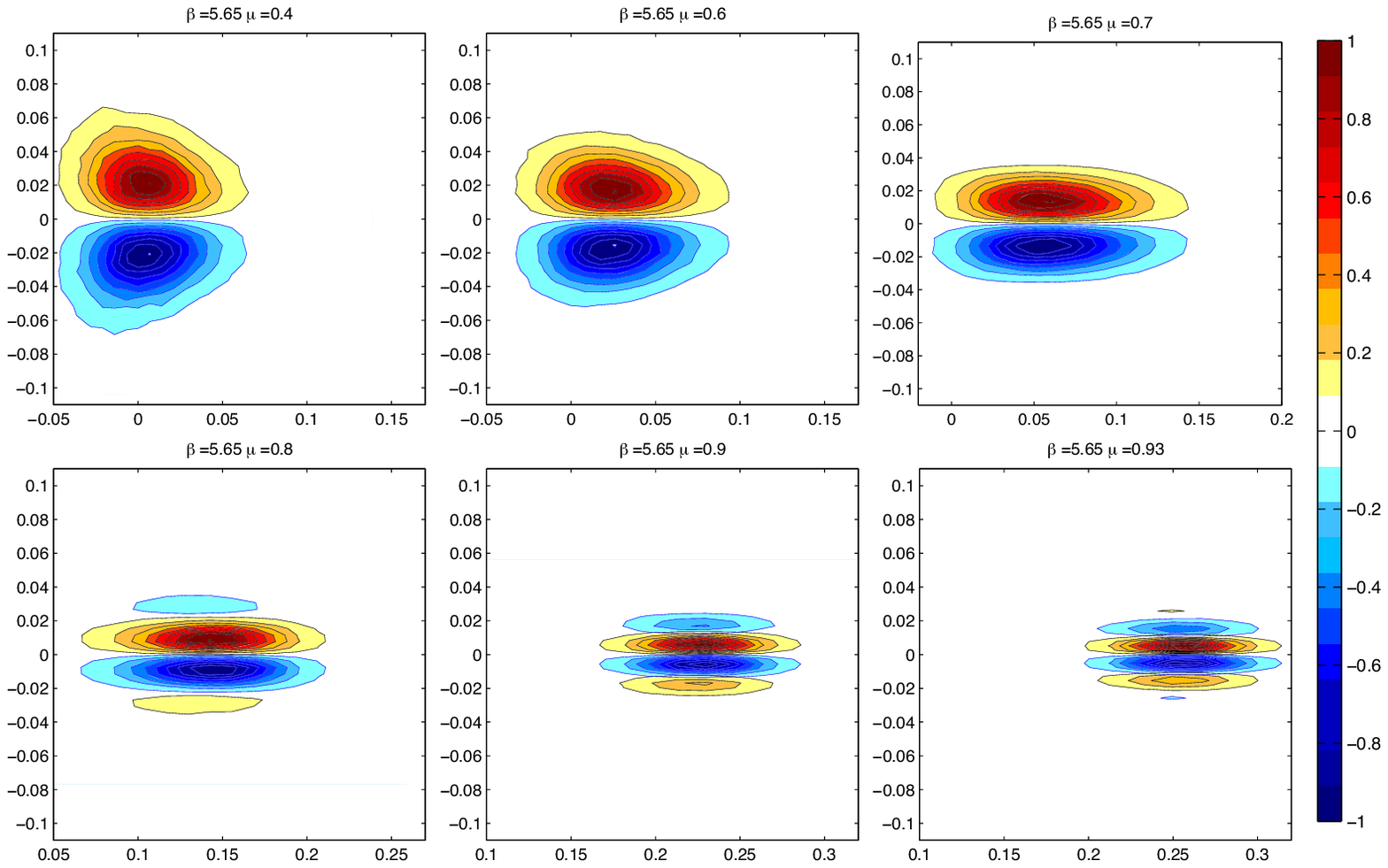}
\caption{Imaginary part of the Polyakov loop `distribution'
$T_{\Delta}(x,y)$ of eq.\ (\ref{eq:Tdistribution})
vs. $\mu$ at $\beta=5.65$ fixed.}
\label{ihismu}
\end{figure*}

\begin{figure*}
\includegraphics{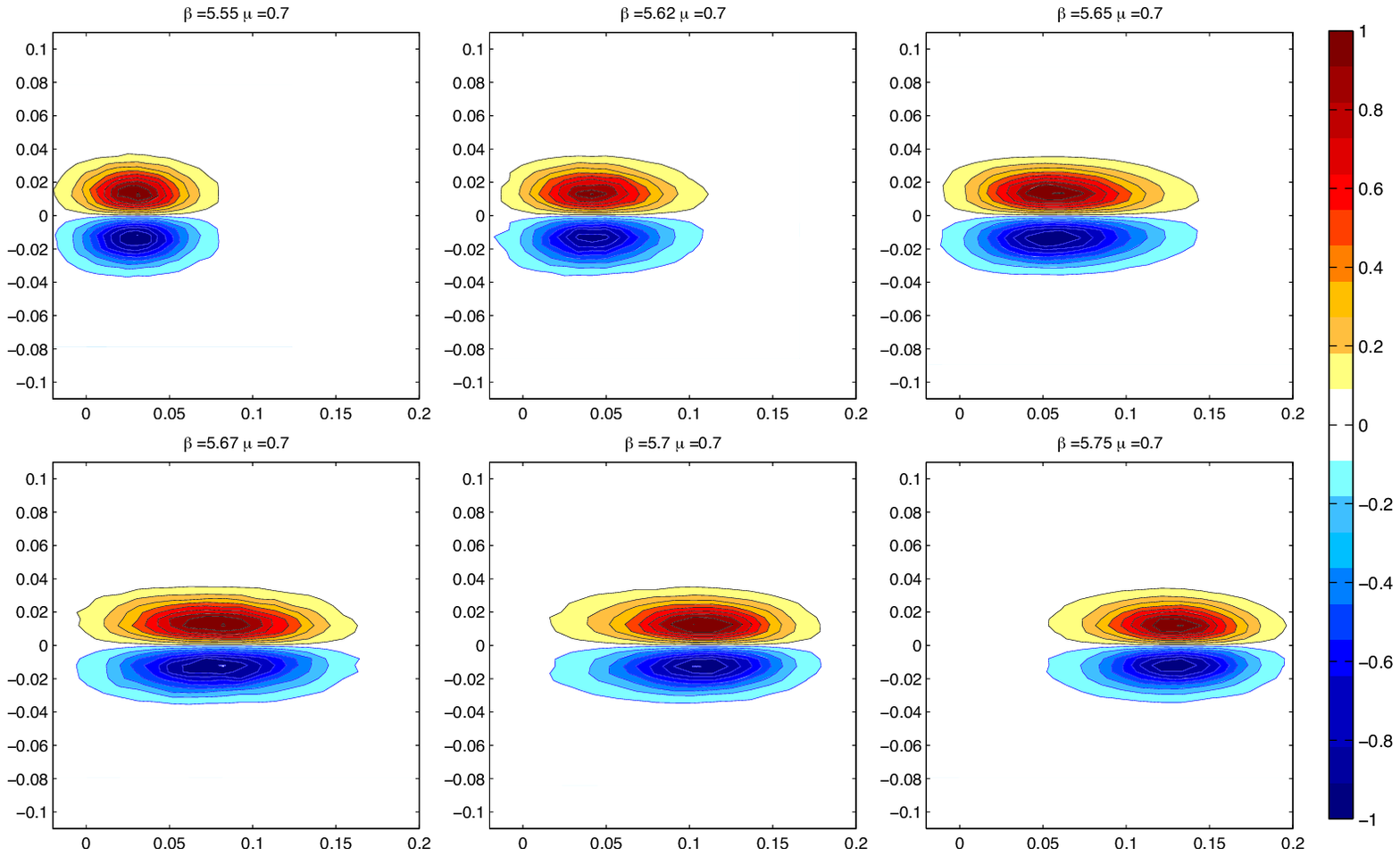}
\caption{Imaginary part of the Polyakov loop `distribution'
$T_{\Delta}(x,y)$ of eq.\ (\ref{eq:Tdistribution})
vs. $\beta$ at $\mu=0.70$ fixed. }
\label{ihisb}
\end{figure*}

Both Fig. \ref{rhismu} and Fig. \ref{rhisb} show that $ReT_\Delta$ 
is to good accuracy even in $y$, as required for the reality of 
$\langle P\rangle$.
In Figs \ref{ihismu} and \ref{ihisb} we show the imaginary parts of the 
`distributions' $T_\Delta$. The qualitative signal of the 
transitions/crossovers is similar to that of $Re T_\Delta$. It should be 
noted that now $ImT_\Delta$ is, to very good precision, odd in
$y$, again in agreement with the reality of $\langle P\rangle$.

\begin{figure} 
\includegraphics[width=6cm,angle=-90]{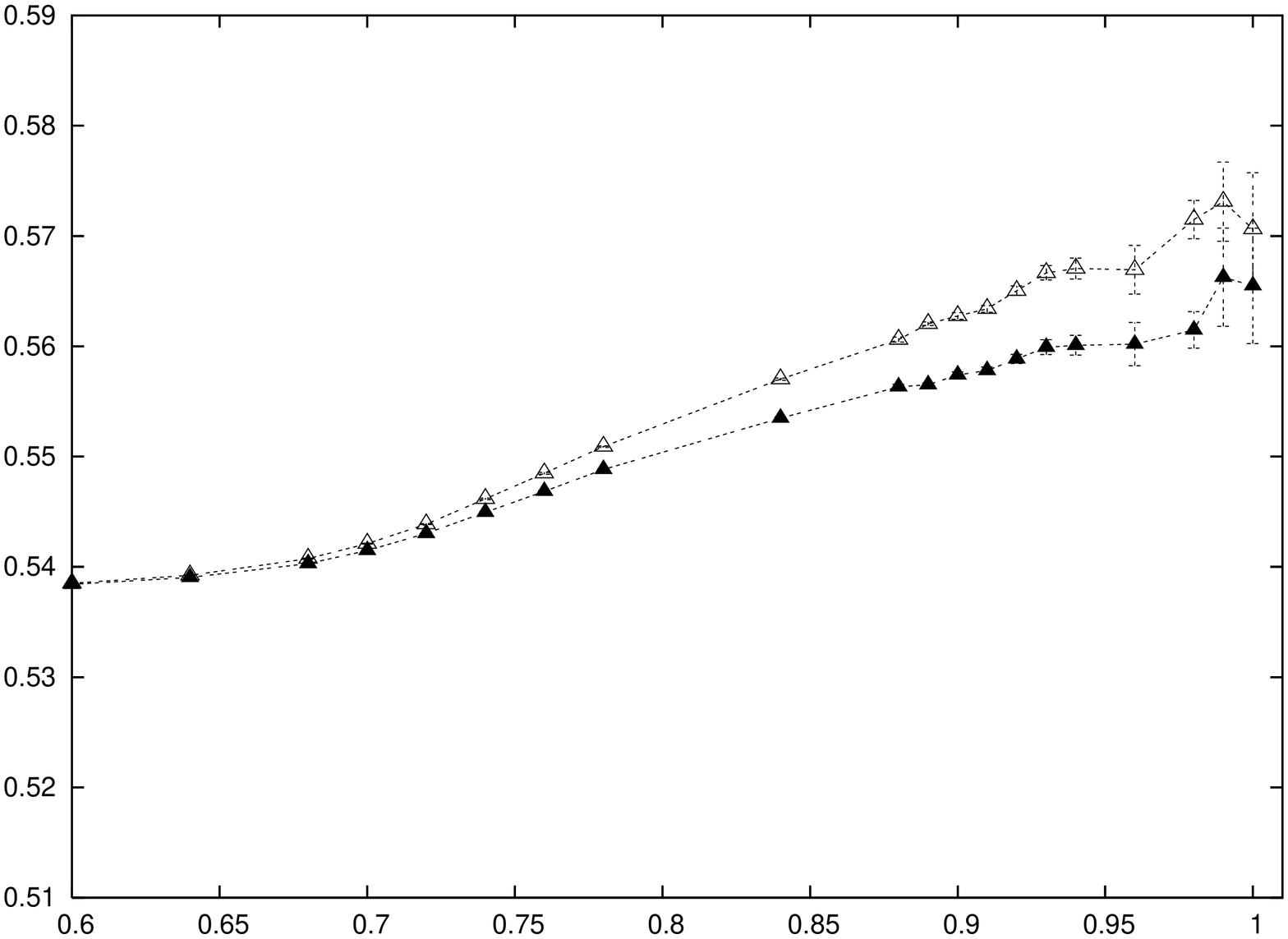}
\caption{Plaquette averages vs. $\mu$ at fixed 
$\beta=5.65$.} 
\label{f.ptsmu} 
\end{figure}

\begin{figure} 
\includegraphics[width=6cm,angle=-90]{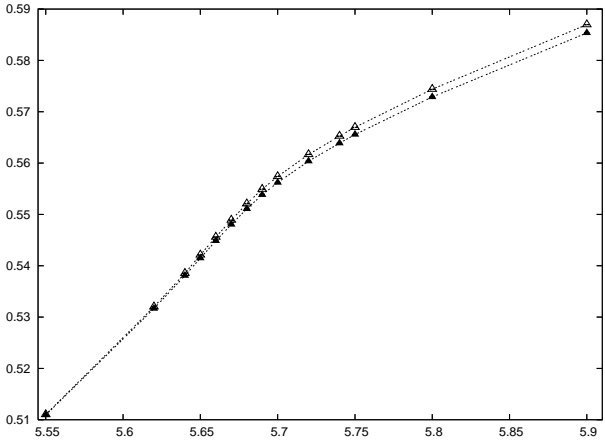}
\caption{Plaquette averages vs. $\beta$ at fixed 
$\mu=0.70$.} 
\label{f.ptsb} 
\end{figure}

\begin{figure} 
\includegraphics[width=6cm,angle=-90]{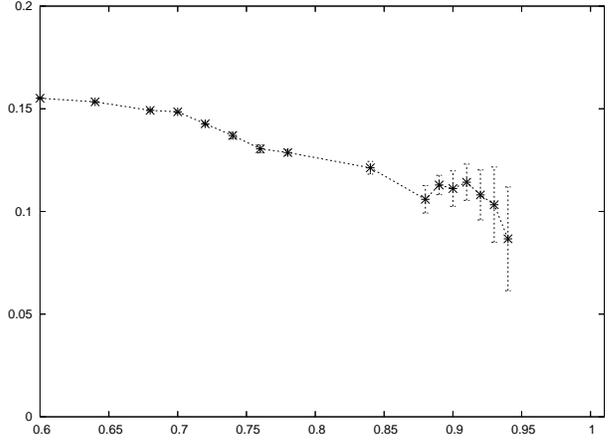}
\caption{Topological susceptibility average vs. 
$\mu$ at fixed
$\beta=5.65$.} 
\label{f.chimu} 
\end{figure}

\begin{figure} 
\includegraphics[width=6cm,angle=-90]{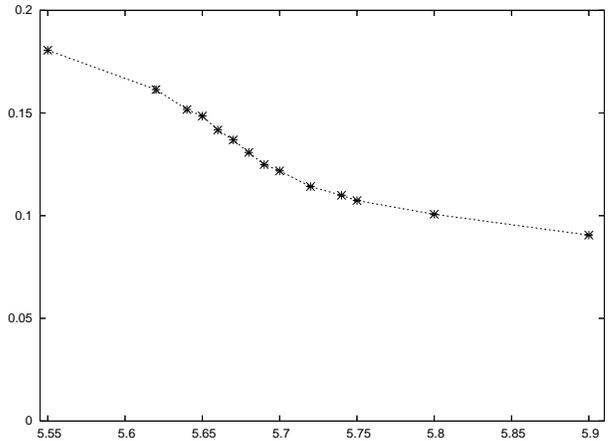}
\caption{Topological susceptibility average vs.
 $\beta$ at fixed
$\mu=0.70$.} 
\label{f.chib} 
\end{figure}

\begin{figure} 
\includegraphics[width=6cm,angle=-90]{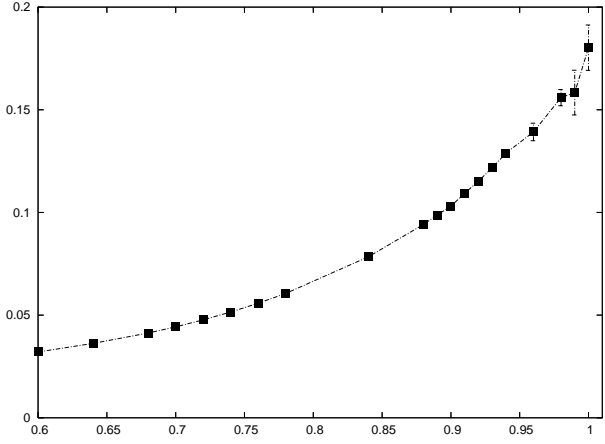}
\caption{Diquark susceptibility average vs. $\mu$ at fixed
$\beta=5.65$.} 
\label{f.qq2mu} 
\end{figure}

\begin{figure} 
\includegraphics[width=6cm,angle=-90]{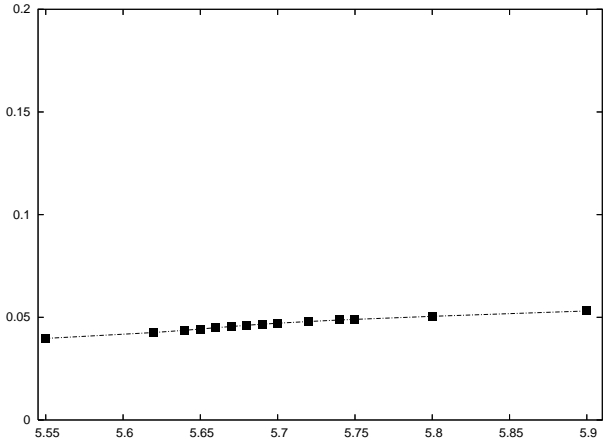}
\caption{Diquark susceptibility average vs. $\beta$ at fixed
$\mu=0.70$.} 
\label{f.qq2b} 
\end{figure}

Polyakov loops and charge density (and their susceptibilities), have been 
the primary quantities used to uncover the phase structure. We also have 
measured plaquette averages (for both temporal and spatial plaquettes), 
the topological charge density (using the improved
field definition) and quark and di-quark correlators (in maximal axial 
gauge). All these quantities also some show peculiar behavior in both 
$\mu$ and $\beta$ which will be exemplified here on two chosen runs, at 
fixed $\beta=5.65$ vs. $\mu$ and at fixed $\mu=0.7$ vs. $\beta$:
In Figs. \ref{f.ptsmu} and \ref{f.ptsb} we present the dependence of the 
plaquette averages on $\mu$ at $\beta=5.65$ and on $\beta$ at $\mu=0.7$, 
respectively. We see here clearly the emergence of a physical energy 
density by the gap developing between the spatial and temporal plaquettes 
with increasing $\mu$ and $\beta$; this corroborates the phase picture 
derived before. In Figs. \ref{f.chimu} and \ref{f.chib} we present for the 
same runs the topological susceptibility whose behavior again is in 
agreement with the previous conclusions since it decreases 
in the region where we
expect deconfining to set in. Finally in Figs. \ref{f.qq2mu} 
and \ref{f.qq2b} we present the dependence on $\mu$ and on $\beta$ of the 
diquark susceptibility obtained by integrating the
diquark-correlators Eq.(\ref{dq}) for $\xi=0.5$; here we 
only show the contribution to this susceptibility from the $\kappa^2$ 
terms. This corresponds to quarks 
showing a (limited) amount of mobility and 
as can be seen from these figures, the susceptibility to this 
order is sensitive to the chemical potential (while the zero-th order 
contribution is dominated by a contact term and is rather flat). The 
strong increase with $\mu$, compared with the rather flat $\beta$ 
dependence may indicate new properties of the matter at high density.


\section{Conclusions}
\label{conclusions}
To obtain analytic informations about our model 
we first analyzed it via the strong coupling expansion; the 
agreement for $\beta\le 5.5 $ and small $\mu$ with the numerical 
simulations should be seen as a validation of the simulation program. But 
our calculations show strong effects at slightly larger $\mu$ , which 
already at $\beta=5.6$ depart considerably from strong coupling estimates; 
this is an indication of a possible phase transition. Next we obtained a 
phase diagram in a mean field approximation, showing the existence of 
three different phases.

The phase structure found  by the numerical simulations for $n_f=3$ 
is shown in Fig. \ref{phtm4}. The signal for the deconfining transition
(or narrow crossover) on the line connecting A and B is rather good and 
it also 
appears that at small $\mu$ (above B) the transition is 
smoothed out in accordance with the expectations from full QCD 
simulations \cite{karrev},\cite{afks}. A second transition at large 
$\mu$ could only be identified tentatively. In this region, the diquark 
susceptibility grows strongly. This region needs further study to reach 
a conclusion, but it is interesting that the general picture shows 
qualitative agreement with the one found in the mean field approximation. 

The algorithm works reasonably well over a wide range of parameters and 
for lattices up to $6^4$ ($8^4$ for $n_f=1$). We obtain large densities for 
temperatures $\sim \frac{1}{2}\,T_c$ or less and  reach ratios 
$\frac{\mu_{phys}}{T} \sim 5$. It appears difficult, 
however, to go to larger lattices and larger $\mu$ with this algorithm and 
one should consider improving it. For the time being, however, these
difficulties precluded us from performing further tests, such as finite 
size analysis, in order to establish unequivocally the character of the 
various transitions. 

The model permits to vary $\mu$, $\kappa$, $\beta$ and $N_\tau$ as independent 
parameters. Also anisotropic lattices can be envisaged. It is therefore 
interesting to extend the study to take advantage of this full 
variability. Also extending the model to  higher orders in $\kappa$ 
can be envisaged. The bookkeeping soon becomes unmanageable, one could 
however consider using statistical ensembles of large loops \cite{mn}.

A related matter is the relation to physical quantities such as 
temperature and masses. In this study we introduced a $T$-dependence by 
varying $\beta$ and tried to avoid the necessity of defining a scale by 
considering only dimensionless ratios such as $\mu_{phys}/T$. This, 
however, has to be taken with a grain of salt: indeed, varying 
$\beta$ also introduces varying finite volume and quark `mass' 
effects. It would be less ambiguous to vary $N_\tau$ if we could reach 
large enough lattices. Alternatively one could consider using a variable 
anisotropy. In a first approximation one could take 
$\gamma_G=\gamma_F=\gamma_{phys}$, such as in the mean field approximation 
in section III.B, but non-perturbative corrections might be large and a 
bona-fide calibration may become necessary \cite{bkns}. All 
renormalization questions, however, are difficult when we need to consider 
the effects of the quarks as introduced in fixed order hopping parameter 
expansion.
 
Concerning the significance of this analysis we can take two points of 
view:

Firstly, we can consider this model for itself, as describing 
`quasi-static charges' interacting via gauge forces and having a
non-trivial phase structure.

Secondly, we can consider this model as an evolved `quenched 
approximation' in the presence of charged matter. Then this study would 
give us information about the modified gluon dynamics of the SU(3) theory 
in this situation. It would then be natural to think of it as providing a 
heavy, dense, charged background for propagation of light quarks and 
calculate light hadron spectra and other hadronic properties under such 
conditions. This could also help fixing a scale controlling the behavior 
of the light matter. We consider pursuing work on this subject. \parm

\begin{acknowledgments}

We thank P. de Forcrand for helpful criticism of an earlier version of 
this paper.
The calculations have been done on the VPP5000 computer at the University of
Karlsruhe and on the PC Cluster at the Physics Department of the 
University of Parma. 

\end{acknowledgments}

\appendix*

\subsection{Strong coupling expansion: some details} 

We first calculate the term 
of order zero, which would vanish trivially without the presence of 
the chemical potential term $C$. The fermion determinant to order 
$\kappa^0$ is 
\be
{\cal Z}_F^{[0]}=\prod_{\vec x} {\rm det} (\one+C{\cal P}_{\vec x})^2\ ,
\end{equation}
where the determinant only refers to the color degrees of freedom. In 
order to evaluate this explicitly we introduce the characters 
$\chi_\sigma$ of the irreducible representations $\sigma$ of $SU(3)$.
In the maximal temporal gauge ${\cal P}_{\vec x}$ is simply given by 
$V_{\vec x}$ and we find
\be
{\cal Z}_F^{[0]}=\prod_{\vec x} \left(1+C\chi_3(V_{\vec x})+C^2 
\chi_{\bar 3}(V_{\vec x})+C^3\right)^2 \, .
\end{equation}
Using the well-known facts (see for instance \cite{itznau,gourdin}) 
\bea
\chi_{\bar 3}\chi_3=\chi_1+\chi_8\ ,\cr
\chi_3\chi_3=\chi_{\bar 3}+\chi_6\ ,\cr
\chi_{\bar 3}\chi_{\bar 3}=\chi_3+\chi_{\bar 6}\ ,
\end{eqnarray}
and defining $D\equiv 1+4C^3+C^6$ this becomes
\bea
{\cal Z}_F^{[0]}&=&D^{N_\sigma}\prod_{\vec x}\Bigl[1+
\frac{2C+3C^4}{D}\chi_3(V_{\vec x})\cr
&+&\frac{3C^2+2C^5}{D}\chi_{\bar 3}(V_{\vec x})
+\frac{1}{D}C^2\chi_6(V_{\vec x})\cr
&+&\frac{1}{D}C^4\chi_{\bar 6}(V_{\vec x})
+\frac{2}{D}C^3 \chi_8(V_{\vec x})\Bigr] \ .
\label{zf0}
\end{eqnarray}
From this it is straightforward to obtain the expectation values 
$  \langle P_{\vec x}\rangle$ and $\langle 
P^{\ast}_{\vec x}\rangle$ to order $0$ as
\be
\langle P\rangle^{[0]}= C^2 \frac{1+\frac{2}{3} C^3} {1+4C^3+C^6}
\end{equation}
and 
\be
\langle P^{\ast}\rangle^{[0]}=C\frac{\frac{2}{3}+C^3}{1+4C^3+C^6}\ .
\end{equation}
The next nontrivial order is $O(\kappa^2)$ in the fermion determinant and 
comes from the Polyakov loops with one excursion to a neighboring site. A 
nonzero result is obtained only by combining it with terms from the 
Yang-Mills action; the lowest nontrivial contribution is therefore 
$O(\kappa^2\beta)$. Concretely we obtain to order $\kappa^2$
\be
\frac {{\cal Z}_F^{[2]}}{{\cal Z}_F^{[0]}}= 
\left(1+2C\kappa^2\sum_{\vec x,i,t,t'} \Tr 
{\cal P}_{\vec x,i,t,t'}\right)\ .
\end{equation}
After integrating over the spatial gauge fields $U$ only terms with 
$t'=t+1$ survive; the integrals occurring are of the form
\be
\int \!\! dU Re \Tr \!\! \left(U_{(\vec x,t)i}
U^\dagger_{(\vec x,t)i}\right)\,
\! \Tr \! \! \left(V_{\vec x} U^\dagger_{(\vec x,t)i}U_{(\vec 
x,t)i}\right)=\frac{1}{6} \Tr V_{\vec x} \ .
\end{equation}
Thus we obtain before the integration over the $V$'s
\be
\int \prod dU{\cal Z}_F^{[2]}={\cal Z}_F^{[0]} \left(1+\sum_{\vec 
x}\beta\hat C\chi_3(V_{\vec x})\right)
\end{equation}
with $\hat C\equiv 2\beta C (N_\tau-1)\kappa^2/3$. To obtain the expectation 
values of the Polyakov loops from this we have to expand the product in 
irreducible characters; we need only the terms involving the representations 
$3$, $\bar 3$, $1$. Using Eq.(\ref{zf0}) we see that we need a few more 
decompositions of $SU(3)$ representations, namely
\bea
\chi_3\chi_6&=&\chi_8+\chi_{10}\cr
\chi_3\chi_{\bar 6}&=&\chi_{\bar 3}+\chi_{\bar {15}}\cr
\chi_3\chi_8&=&\chi_3+\chi_{\bar 6}+\chi_{15}\ .
\end{eqnarray}
Since the expectation values are normalized by the partition function, as 
usual only connected contributions occur; thus the results for 
$\langle P\rangle$ and 
$\langle P^\ast \rangle$ to order $\kappa^2$ are
\begin{multline}
\langle P\rangle^{[2]}\equiv C^2 \frac{1+\frac{2}{3} C^3}{1+4C^3+C^6}
\Biggl[1+\cr\frac{2\beta\kappa^2(N_\tau-1)}{3}
\frac{2+3C^2+6C^6}{(1+4C^3+C^6)(3+2C^3)}\Biggr]
\end{multline}
and
\begin{multline}
\langle P^\ast \rangle^{[2]} \equiv 
C \frac{\frac{2}{3}+C^3}{1+4C^3+C^6} 
\Biggl[1+\cr\frac{2\beta\kappa^2(N_\tau-1)}{3}
\frac{(1+C^3)^4+7C^6}{(1+4C^3+C^6)(2+3C^3)}\Biggr]\ .
\end{multline}
We note the leading behavior for small $C$:
\be
P^{[2]}\sim C^2\left(1+\frac{4}{9}\beta\kappa^2(N_\tau-1)\right)
\end{equation}
and
\be
P^{\ast [2]}\sim \frac{2}{3}C
\left(1+\frac{1}{3}\beta\kappa^2(N_\tau-1)\right) \ .
\end{equation}

\subsection{Mean Field: some details}

We first compute the Faddeev-Popov determinant $J(v)$ for the Polyakov 
gauge, which can be computed as the Jacobian for the transformation from 
the maximal temporal to the Polyakov gauge.

The reduced Haar measure for the conjugacy classes $[U]$ of $SU(N)$ is 
given by \cite{weyl} 
\be d[U]= \frac {1}{\cal{N}} \prod_{i<j} 
\sin^2\left(\frac{\phi_i-\phi_j}{2}\right) d\phi_1\ldots d\phi_{N-1}\ , 
\end{equation} 
where $\cal{N}$ is a normalization constant; this would be the appropriate 
measure for the temporal gauge field in the unfixed links of the maximal 
temporal gauge. We are instead spreading the field uniformly over $N_\tau$ 
links such that we want to integrate over $V\in SU(N)$ with 
$V^{N_\tau}=U$, so we want to write \be d[U]=J(V) d[V]\ , \end{equation} 
where $J(V)$ is now the `quotient' of the Haar measures for $V^{N_\tau}$ 
and $U$, i.e. 
\be J(V)= \prod_{i<j} 
\frac{\sin^2\left(\frac{N_\tau(\phi_i-\phi_j)}{2}\right)} 
{\sin^2\left(\frac{\phi_i-\phi_j}{2}\right)}\ . 
\end{equation} 
So we have to integrate the homogeneous temporal gauge fields with the 
measure 
\be 
d[V]=\prod_{i<j}\sin^2\left(\frac{N_\tau(\phi_i-\phi_j)}{2}\right) 
\prod_{k=1}^{N-1}d\phi_k\ . 
\end{equation}
The range of integration is the interval $[-\pi,\pi)$ for each $\phi_i$; 
this means of course that $V^{N_\tau}$ covers the group $SU(N)$ $N_\tau$ 
times; this `over-counting' is necessary, since otherwise the integration of 
functions of $V$ would involve some completely arbitrary choice of the 
`$N_\tau$th' root.

We now proceed in the standard fashion to produce the mean field theory 
as a saddle point approximation (see for instance \cite{dz,mr}) for the 
partition function: first the integrals over the group $SU(N)$ are 
replaced by integrals over the embedding matrix space $M_{N,N}(\C)$ by 
inserting the identities 
\bea 
1&=&\int\limits_{M_{N,N}}du 
\delta(U-u)\cr 
&=&c\int\limits_{M_{N,N}} \!\! dM \int\limits_{M_{N,N}} \!\! du\, 
\exp\left[i Re \Tr M^\dagger(U-u)\right] \ \ 
\end{eqnarray} 
for each spatial link and similarly for $V$, introducing the matrix 
valued fields $v$ and $K$ for each temporal link. The group integrals for 
the different links are then decoupled and reduce to the one-link integrals 
\be
\int dU \exp(Re \Tr M^\dagger U)
\end{equation}
and
\be
\int dV J(V) \exp(Re \Tr K^\dagger V)\ .
\end{equation}
Carrying out the integrals over the gauge field, using these definitions, 
the partition function reduces to an integral over the matrix valued 
fields $u,v,M,K$ with an effective action $\tilde S(u,v,M,K)$.  This 
integral is suitable for a saddle point approximation. By symmetry there 
must be a translation invariant extremal of $\tilde S$. For the matrix 
valued fields we furthermore make the ansatz that they are multiples of 
the identity; by slightly abusive notation
\bea 
&u&=u\one,\ \ v=(v_1+iv_2)\one,\cr &M&=(m_1+im_2)\one, \ \ K=(k_1+ik_2)\one\ . 
\end{eqnarray}
We anticipated here already that $u$ will be real. Using this ansatz 
and introducing a single asymmetry parameter $\gamma=\gamma_G=\gamma_F$, 
as discussed in Section \ref{analytic}, the action per site $\tilde s$ 
becomes
\bea
-&\tilde s& = 3\frac{\beta}{\gamma} u^4 + 
3 \beta\gamma u^2(v_1^2+v_2^2)\cr 
&+& 6C(v_1+iv_2)^{N_\tau}
\left(N_\tau^{-1}+ 3(N_\tau-1)\kappa^2u^2\right) \cr 
&+&3 \ln \zeta (im_1)+\ln \eta(ik_1,ik_2)\cr 
&-&3i(k_1v_1+k_2v_2)-9im_1u
\label{tildes}
\end{eqnarray}
where the functions $\zeta$ and $\eta$ are defined for arbitrary 
complex arguments $a, b_1, b_2$ as
\be
\zeta(a)\equiv \int d[U] \exp(a Re \Tr U) 
\end{equation}
and 
\be 
\eta(b_1,b_2)\equiv \int d[V] J(V)\exp(b_1 Re \Tr V+ 
b_2\ Im \Tr V)\ 
\end{equation}
For the group $SU(3)$ we write the functions $\zeta$ and $\eta$ in 
more explicit form:
\begin{multline}
\zeta(a)=\int_{-\pi}^\pi d\phi_1\int_{-\pi}^\pi d\phi_2 
\rho_1(\phi_1,\phi_2)\\
\times \exp\left[a(\cos\phi_1+\cos\phi_2+\cos(\phi_1+\phi_2)\right] 
\end{multline}
and
\begin{align}
\eta(b_1,b_2)&= \int_{-\pi}^\pi d\phi_1\int_{-\pi}^\pi d\phi_2
\rho_{N_\tau}(\phi_1,\phi_2) \nn \\
&\times\exp\left[b_1(\cos\phi_1+\cos\phi_2+\cos(\phi_1+\phi_2)\right] \nn \\
&\times \exp\left[b_2(\sin\phi_1+\sin\phi_2-\sin(\phi_1+\phi_2)\right] \,,
\end{align}
with
\bea 
\rho_k(\phi_1,\phi_2)&=&\sin^2 \left(\frac{k(\phi_1-\phi_2)}{2}\right)
\sin^2 \left(\frac{k(\phi_1+2\phi_2)}{2}\right)\cr
&\times&\sin^2 \left(\frac{k(\phi_2+2\phi_1)}{2}\right) 
\end{eqnarray}

When searching for a saddle point we have to allow all 
parameters to be complex. The saddle point equations, requiring 
stationarity of $\tilde s$ with respect 
to 
$u,v_1,v_2,a=im_1,b_1=ik_1,b_2=ik_2$ are
\bea
a&=&\frac{4}{3}\frac{\beta}{\gamma} u^3+\frac{2}{3}\beta\gamma 
u(v_1^2+v_2^2)\cr &+&
4 C\kappa^2(N_\tau-1)u(v_1+iv_2)^{N_\tau}\ , \cr
b_1&=&2\frac{\beta}{\gamma} u^2v_1\cr
&+&2C(v_1+iv_2)^{N_\tau-1}(1+3N_\tau(N_\tau-1)\kappa^2u^2)
\ , \cr
b_2&=&2\frac{\beta}{\gamma} u^2v_2\cr
&+&2iC(v_1+iv_2)^{N_\tau-1}(1+3N_\tau(N_\tau-1)
\kappa^2u^2)
\ ,\cr
u_{\ }&=&\frac{1}{3} \frac{\rm d}{{\rm d}a} \ln\zeta(a)\ ,\cr
v_1&=&\frac{1}{3} \frac{\partial}{\partial b_1} \ln\eta(b_1,b_2)\ , \cr
v_2&=&\frac{1}{3} \frac{\partial}{\partial b_2} \ln\eta(b_1,b_2)\ .
\end{eqnarray}
The system of equations is of the form of a fixed point condition 
and is solved by iteration. There is always a trivial fixed point 
\be
u=v_1=v_2=a=a_1=b_1=0\ .
\end{equation}
In general if there is more than one fixed point (which may be reached by 
choosing different starting points for the iteration). It turns out that 
all the fixed points satisfy $a=im_1$ real, $b_1=ik_1$ purely imaginary, 
$v_2=0$ and $u,v_1$ real; note that $v_2=0$ is consistent with these 
equations because of the
symmetry
\be
\eta(b_1,b_2)=\eta(b_1,-b_2)\ ,
\end{equation}
which follows from the unimodularity ($d[U]=d[U^\dagger])$.
 
With our sign convention one always has to choose the fixed point leading 
to the highest value of the free energy density $f=\tilde s$ for the 
parameters chosen. This leads to discontinuities in the first derivative, 
typical for first order phase transitions, and finally to the phase 
diagram shown in Fig.\ref{mf}.

\end{document}